\documentclass[journal]{IEEEtran}
\ifCLASSINFOpdf
  \usepackage[pdftex]{graphicx}
\else
  \usepackage[dvips]{graphicx}
\fi
\usepackage{amsmath}
\usepackage{caption}
\usepackage{amssymb}
\usepackage{bm}
\usepackage{upgreek}
\usepackage{tabularx}
\usepackage{float}
\usepackage{xcolor}
\usepackage[normalem]{ulem}
\usepackage{enumitem}

\newcommand{\KQ}[1]{\textcolor{magenta}{#1}}

\usepackage{booktabs} 
\usepackage{algorithm} 
\usepackage{algorithmic}
\usepackage{multirow}
\usepackage{svg}
\usepackage[switch, pagewise]{lineno}

\begin{document}
%
\title{$\boldsymbol{\gamma}$-Net: Superresolving SAR Tomographic Inversion via Deep Learning}
%
%
%

\author{Kun~Qian,
        Yuanyuan~Wang,~\IEEEmembership{Member,~IEEE,}
        Yilei~Shi, ~\IEEEmembership{Member,~IEEE,}
        and~Xiao~Xiang~Zhu,~\IEEEmembership{Fellow,~IEEE}
\thanks{This work is supported by China Scholarship Council with the grant No.201908080038, by the European Research Council (ERC)  with the grant agreement No. [ERC-2016-StG-714087], Acronym: So2Sat, by the Helmholtz Association through the Framework of Helmholtz AI (grant  number:  ZT-I-PF-5-01) - Local Unit ``Munich Unit @Aeronautics, Space and Transport (MASTr)'', and Helmholtz Excellent Professorship ``Data Science in Earth Observation - Big Data Fusion for Urban Research''(grant number: W2-W3-100) and by the German Federal Ministry of Education and Research (BMBF) in the framework of the international future AI lab "AI4EO -- Artificial Intelligence for Earth Observation: Reasoning, Uncertainties, Ethics and Beyond" (grant number: 01DD20001).

K. Qian, Y. Wang and X. Zhu are with the Data Science in Earth Observation, Technical University of Munich, Munich, Germany. Y. Wang and X. Zhu are also with the Department of EO Data Science, Remote Sensing Technology Institute, German Aerospace Center, Oberpfaffenhofen, Germany.}
\thanks{Y. Shi is with the Chair of Remote Sensing Technology, Technical University of Munich, Munich, Germany.}
}

%
%

\markboth{Journal of \LaTeX\ Class Files,~Vol.~14, No.~8, August~2015}%
{Shell \MakeLowercase{\textit{et al.}}: Bare Demo of IEEEtran.cls for IEEE Journals}
%



\maketitle

\begin{abstract}
\KQ{(This work has been submitted to the IEEE for possible publication. Copyright may be transferred without notice, after which this version may no longer be accessible.)}
Synthetic aperture radar tomography (TomoSAR) has been extensively employed in 3-D reconstruction in dense urban areas using high-resolution SAR acquisitions. Compressive sensing (CS)-based algorithms are generally considered as the state of the art in super-resolving TomoSAR, in particular in the single look case. This superior performance comes at the cost of extra computational burdens, because of the sparse reconstruction, which cannot be solved analytically and we need to employ computationally expensive iterative solvers. In this paper, we propose a novel deep learning-based super-resolving TomoSAR inversion approach, $\boldsymbol{\gamma}$-Net, to tackle this challenge. $\boldsymbol{\gamma}$-Net adopts advanced complex-valued learned iterative shrinkage thresholding algorithm (CV-LISTA) to mimic the iterative optimization step in sparse reconstruction. Simulations show the height estimate from a well-trained $\boldsymbol{\gamma}$-Net approaches the Cram\'er-Rao lower bound while improving the computational efficiency by 1 to 2 orders of magnitude comparing to the first-order CS-based methods. It also shows no degradation in the super-resolution power comparing to the state-of-the-art second-order TomoSAR solvers, which are much more computationally expensive than the first-order methods. Specifically, $\boldsymbol{\gamma}$-Net reaches more than $90\%$ detection rate in moderate super-resolving cases at 25 measurements at 6dB SNR. Moreover, simulation at limited baselines demonstrates that the proposed algorithm outperforms the second-order CS-based method by a fair margin. Test on real TerraSAR-X data with just 6 interferograms also shows high-quality 3-D reconstruction with high-density detected double scatterers.


\end{abstract}


%
\IEEEpeerreviewmaketitle

\section{Introduction}
%
%
%
%
\IEEEPARstart{S}{ynthetic} aperture radar (SAR) tomography (TomoSAR) \cite{tomosar} has been widely employed for large-scale 3-D urban mapping. It utilizes a stack of SAR acquisitions to reconstruct the reflectivity profile $\boldsymbol{\gamma}$ along the elevation direction for every azimuth-range pixel. In urban areas, there are usually only a few significant scatterers overlaid in a resolution cell along the elevation direction. Based on this fact, compressive sensing (CS)-based sparse reconstruction algorithms \cite{CS1,CS2, CS3} were introduced to TomoSAR inversion so that we can best unleash the potential of high-resolution SAR data like TerraSAR-X in urban areas. \cite{cs_tomosar1, zhu_very_2009} presented the first simulation of CS TomoSAR, \cite{zhu_very_2009} presented the first real data example and \cite{Zhu2010Tomographic} proved the super-resolution power of CS for TomoSAR inversion. In recent years, different CS-based methods for solving TomoSAR inversion have been extensively studied, such as SL1MMER \cite{Zhu2012Super-Resolution}, TSVD-based CS \cite{cs_tomosar2} and ADMM-based $L_1$ algorithm \cite{cs_tomosar4}. These CS-based algorithms show superiority in super-resolution capability as well as elevation estimate accuracy over the conventional $L_2$ regularization methods. However, CS-based algorithms usually suffer from high computational expense and are more challenging to be extended to large-scale processing. An efficient approach was proposed in \cite{Wang2014Efficient} to address this issue, which is an integration of persistent scatterer interferometry (PSI) and "\textit{SL1MMER}". This approach speeds up the processing by pre-classifying the pixels and reducing the percentage of pixels that require SL1MMER for sparse reconstruction. Nevertheless, it did not boost the TomoSAR inversion fundamentally. The same authors of \cite{Wang2014Efficient} also proposed a data-driven method \cite{BSS}, which is based on the CAESAR algorithm \cite{PCA}. It applies kernel principle component analysis (KPCA) to separate the contribution of individual scatterers before inversion, thus reducing the computational cost logarithmically. Although these algorithms brings a perspective of data-driven approaches in TomoSAR, they still do not strictly solve the SAR tomographic inversion. Their super-resolution capability are also not investigated. Therefore, there has not been a fully data-driven TomoSAR algorithm to date. Hence, we would like to explore the potential of modern data science algorithm such as deep learning, for TomoSAR in this paper.

\subsection{Related work}
Recently, deep learning has rapidly developed and been extensively applied in various fields of remote sensing \cite{zhu2017dl}, including SAR data processing \cite{zhu2021dl4sar}, thanks to its strong learning power. In particular, a deep neural network can act as an effective nonlinear function and is capable of representing many complicated mathematical models including the CS problems \cite{deep_cs}. Several recent studies \cite{RMIST-Net} \cite{CSR-Net} \cite{S_LISTA} have documented the application of deep neural networks in solving sparse reconstruction related problems in signal processing and remote sensing, i.e. 3-D millimeter-wave sparse imaging and 3-D microwave reconstruction. Motivated by this fact, community started to investigate TomoSAR inversion algorithms based on deep learning and their application since a few years ago. \cite{dl_tomosar1} proposed a method to utilize neural networks to detect single scatterer and estimate the corresponding elevation. In \cite{dl_tomosar1}, TomoSAR inversion was treated as a typical classification problem to detect single scatterer with the classes indicating all the possible discretized positions within the elevation extension of the illuminated scene. Because of its problem formulation, this method cannot be employed in true SAR 3-D imaging, i.e., layover separation. An efficient line spectral estimation algorithm based on deep neural networks was proposed in \cite{gao2018fast} and applied to tackle the TomoSAR inversion. Experiment results in \cite{gao2018fast} showed that the method can separate overlaid scatterers and achieves moderate reconstruction performance, whereas the super-resolution power of the proposed method was not systematically analyzed. More recently, a novel super-resolving TomoSAR imaging framework based on CS and deep neural networks was proposed in \cite{MIMO}. It employed CS-based algorithms for preliminary reconstruction and split the elevation range to several subregions with spatial filters. Then a group of deep neural networks based regression models were trained and applied to each subregion to achieve final super-resolution reconstruction. This method was shown to have unprecedented super-resolution capability. However, the drawback of the proposed algorithm is also obvious. First, the computational complexity of the proposed method is of same order of magnitude to other CS-based algorithms, although the authors increased the sampling distance between two neighbor discrete grid point. The second drawback is that the strong super-resolution power is attributed to adequate training samples, whereas it is arduous to simulate data that imitate the real scattering scenario of high fidelity. Hence, strong overfitting to the training data is expected. 



\subsection{Contribution of this paper}
The aim of this paper is to introduce a computationally efficient and generic TomoSAR algorithm based on deep learning and provide a systematic analysis of its super-resolution power. To this end, we propose a deep learning based approach to address super-resolving TomoSAR inversion. We unroll iterative shrinkage thresholding algorithm (ISTA) as a complex-valued feedforward neural network with side-connection, named as $\boldsymbol{\gamma}$-Net. $\boldsymbol{\gamma}$-Net could be trained using data simulated by spatial baselines of given stacked SAR acquisitions. Once well-trained, $\boldsymbol{\gamma}$-Net can be directly used for further inference. The main contributions of this paper are listed as follows:

\begin{enumerate}
    \item We are the first to introduce a deep unfolded neural network called $\boldsymbol{\gamma}$-Net to solve super-resolving TomoSAR inversion. We improved the conventional soft-thresholding function by the piecewise linear function to mitigate the loss of information caused by the learning architecture. Simulations indicate that the piecewise linear function contributes to improving the convergence rate and reducing reconstruction error.
    \item We are the first to perform a systematic evaluation of a deep learning-based TomoSAR inversion algorithm. We investigated the generalization ability w.r.t amplitude ratio, phase difference of the interfering scatterers, super-resolution power and elevation estimation accuracy of the proposed algorithm, i.e. $\boldsymbol{\gamma}$-Net. Experiments demonstrate that $\boldsymbol{\gamma}$-Net not only approaches almost the same performance in nominal condition comparing to the state of the art but also outperforms the state of the art at limited number of measurements.
    \item We carried out rigorous analysis of algorithm complexity and proved that $\boldsymbol{\gamma}$-Net improves the computational efficiency by 1 to 2 orders of magnitude comparing to first-order CS-based methods and shows no degradation in super-resolution power, nor in estimation accuracy comparing to second-order CS-based methods, which are much more computationally expensive than first-order methods. Further time consumption comparison to the second-order CS-based method establishes the superiority of $\boldsymbol{\gamma}$-Net in computational efficiency and evidences that $\boldsymbol{\gamma}$-Net is able to realize large-scale super-resolving TomoSAR processing, whereas second-order CS-based solvers can only be applied to small laboratory samples.
\end{enumerate}
The remainder of the paper is outlined as follows: In section II, the TomoSAR imaging model and inversion are briefly introduced. Section III provides an overview on the formulation of $\boldsymbol{\gamma}$-Net. Results of systematic evaluation, using simulated and real data, are presented and discussed in section IV and V. Finally, the conclusion of this paper is drawn in section VI.

\section{SAR imaging model and problem formulation}
\begin{figure}[H]
    \centering
    \includegraphics[width=0.49\textwidth]{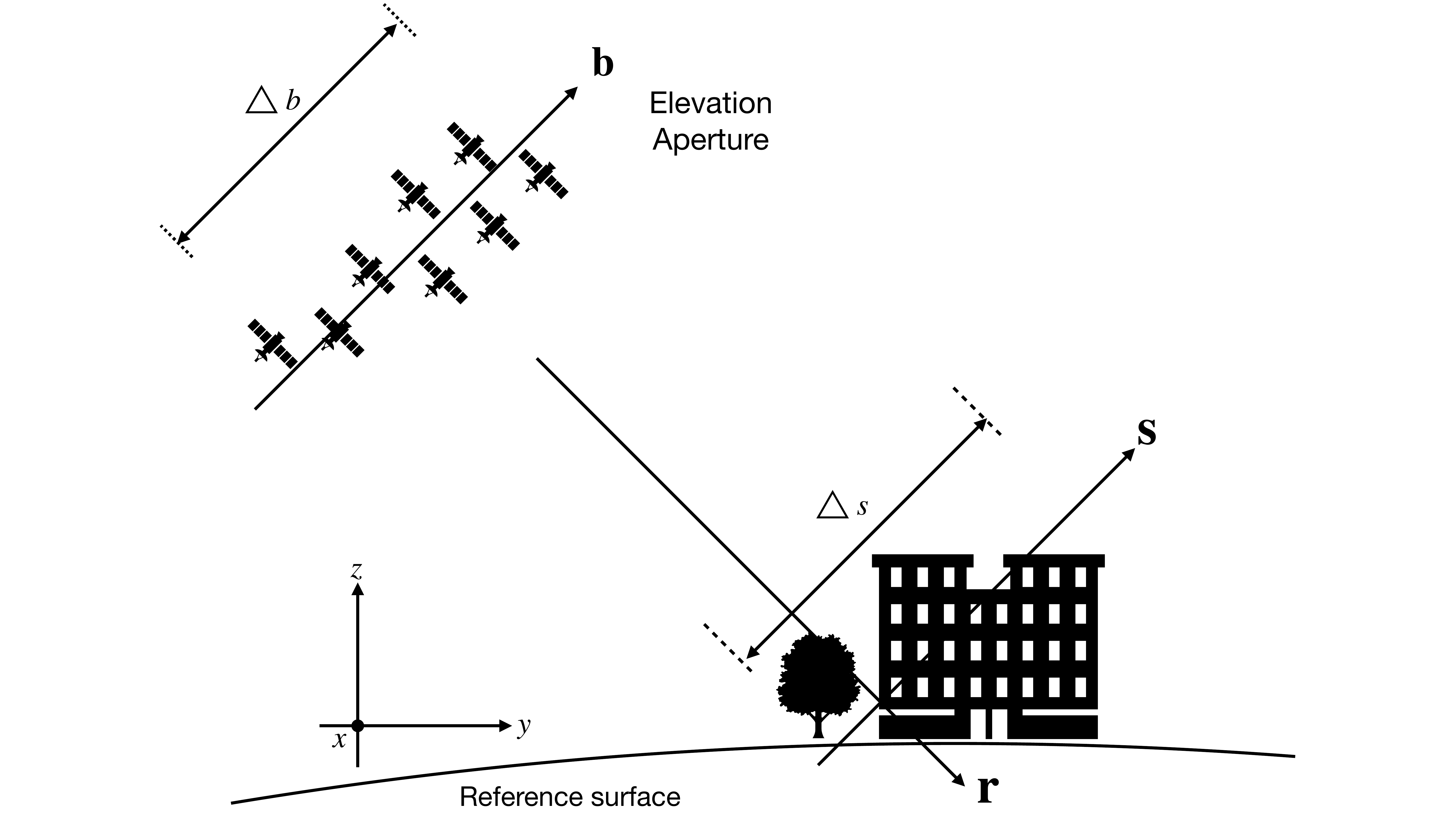}
    \caption{The SAR imaging geometry. The elevation synthetic aperture is built up by SAR data acquired from slightly different viewing angles. Flight direction is orthogonal into the plane.}
    \label{fig:tomosar}
\end{figure}
Firstly, we would like to introduce the TomoSAR imaging model (see Fig. \ref{fig:tomosar}). For a single SAR acquisition at aperture position $b_n$, the complex-valued measurement at an azimuth-range pixel for the $n$th acquisition is the integral of the reflected signal along the elevation direction and can be expressed as follows:
\begin{equation}
    g_{n}=\int_{\Delta s} \gamma(s) \exp \left(-j 2 \pi \xi_{n} s\right) d s, \quad n=1, \ldots, N
\end{equation}
where $\gamma(s)$ depicts the reflectivity profiles along the elevation direction. $\xi_{n}=-2 b_{n} /(\lambda r)$ denotes the elevation frequency.  $\lambda$ and $r$ refer to the wavelength and the range, respectively. By discretizing the reflectivity profile along the elevation direction $s$, the approximated system model reads 
\begin{equation}
    g_{n} \approx \delta s \cdot \sum_{l=1}^{L} \gamma\left(s_{l}\right) \exp \left(-j 2 \pi \xi_{n} s_{l}\right), \quad n=1, \ldots, N
\end{equation}
where $L$ is the number of the discrete elevation indices and $\delta s=\Delta s /(L-1)$ is the sampling distance with $\Delta s$ depicting the whole extent of the reflectivity profile along the elevation direction. In the presence of noise $\boldsymbol{\varepsilon}$, the discrete TomoSAR imaging model can be expressed as follows:
\begin{equation}\label{imaging model}
    \mathbf{g}=\mathbf{R} \boldsymbol{\gamma} + \boldsymbol{\varepsilon},
\end{equation}
where $\mathbf{g} \in \mathbb{C} ^{N \times 1}$ is the complex-valued SAR measurement vector, $\mathbf{R} \in \mathbb{C}^{N \times L}$ is the steering matrix with $R_{n l}=\exp \left(-j 2 \pi \xi_{n} s_{l}\right)$, and $\boldsymbol{\gamma} \in \mathbb{C}^{L \times 1}$ denotes the discrete reflectivity profile vector. TomoSAR inversion is aimed at retrieving the reflectivity profile for each range-azimuth cell, then estimating the corresponding scatterering parameters such as the number of scatterers and their elevation and reflectivity. 

For TomoSAR reconstruction in urban areas, it is shown in \cite{Zhu2010Tomographic} that there are rarely more than a few (0-4) scatterers overlaid along the elevation direction in each resolution unit, namely, the reflected signal along the elevation direction is sufficiently sparse. The ideal sparse solution of $\boldsymbol{\gamma}$ is obtained by solving (\ref{imaging model}) with the $L_0$-norm regularization, which is, however, a NP-hard problem. For our application, it is shown in \cite{CS1, CS2, CS3} \cite{Zhu2010Tomographic} that the $L_0$-norm minimization can be approximate by the $L_1$-norm minimization, which can be expressed as follows:
\begin{equation} \label{gamma estimate}
    \hat{\boldsymbol{\gamma}}=\arg \min _{\boldsymbol{\gamma}}\left\{\|\mathbf{g}-\mathbf{R} \boldsymbol{\gamma}\|_{2}^{2}+\lambda \|\boldsymbol{\gamma}\|_{1}\right\},
\end{equation}
where $\lambda$ is a regularization parameter balancing the sparsity and data-fitting terms. It should be adjusted according to the noise level as well as the desired sparsity level. The choice of a proper $\lambda$ is described in great detail in \cite{BPDN}. The $L_2$-$L_1$ mixed norm minimization (\ref{gamma estimate}) is also known as basis pursuit denoising (BPDN) \cite{BPDN} and can be formulated as least absolute shrinkage and selection operator (LASSO) in some condition. Conventional solvers for (\ref{gamma estimate}) are either first- or second-order methods. First-order methods are typically based on linear approximations, such as ISTA \cite{ISTA}, ADMM \cite{admm}. An example for the second-order methods is the primal-dual interior-point method (PDIPM) \cite{PDIPM}. Second-order methods often suffer from high computational cost, thus impeding their application in large-scale processing.

\section{TomoSAR inversion via $\boldsymbol{\gamma}$-Net}
\subsection{Background of complex-valued ISTA}
ISTA \cite{ISTA} is a popular method to solve the $L_2$-$L_1$ mix norm minimization. Each iteration of ISTA is defined by 
\begin{align}
    \hat{\boldsymbol{\gamma}}_i = \eta_{st} (\hat{\boldsymbol{\gamma}}_{i-1} + \beta \mathbf{R}^H \mathbf{b}_{i-1}, \theta_i) \\ \nonumber
    \mathrm{with} \quad \mathbf{b}_i = \mathbf{g - R} \hat{\boldsymbol{\gamma}}_i,
\end{align}
where $\hat{\boldsymbol{\gamma}}_0 = \mathbf{0}$, $\beta$ is the stepsize, $\eta_{st}$ is the soft-thresholding function applied to each element of $\hat{\boldsymbol{\gamma}}_i$, and $\theta$ is the threshold in the soft-thresholding function. The complex-valued version of the soft-thresholding function $\eta_{st}$ is defined by
\begin{equation}
    \eta_{st}(\hat{\boldsymbol{\gamma}}_i, \theta_i) = e^{j \cdot \angle(\hat{\boldsymbol{\gamma}}_i)}\mathrm{max}(|\hat{\boldsymbol{\gamma}}_i|-\theta_i, 0).
\end{equation}
where $j$ is the imaginary number. In each iteration of ISTA, the estimate is firstly optimized via gradient decent and then the soft-thresholding function is applied to prune the elements with small magnitude, thus promoting the sparsity of the final estimate.

\subsection{Complex-valued LISTA formulation for TomoSAR}
We can rewrite equation (5) as the following form: 
\begin{equation} 
    \hat{\boldsymbol{\gamma}}_{i}=\eta_{s t}\left\{\mathbf{W}_{1}^{i} \mathbf{g}+\mathbf{W}_{2}^{i} \hat{\boldsymbol{\gamma}}_{i-1}, \theta_{i}\right\},
    \label{eq:LISTA}
\end{equation}
where $\mathbf{W}_{\mathbf{1}}^{\boldsymbol{i}}=\beta \mathbf{R}^{H}$ and $\mathbf{W}_{2}^{i}=\mathbf{I}-\beta \mathbf{R}^{\mathrm{H}} \mathbf{R}$.

\begin{figure}[h]
    \centering
    \includegraphics[width=0.49\textwidth]{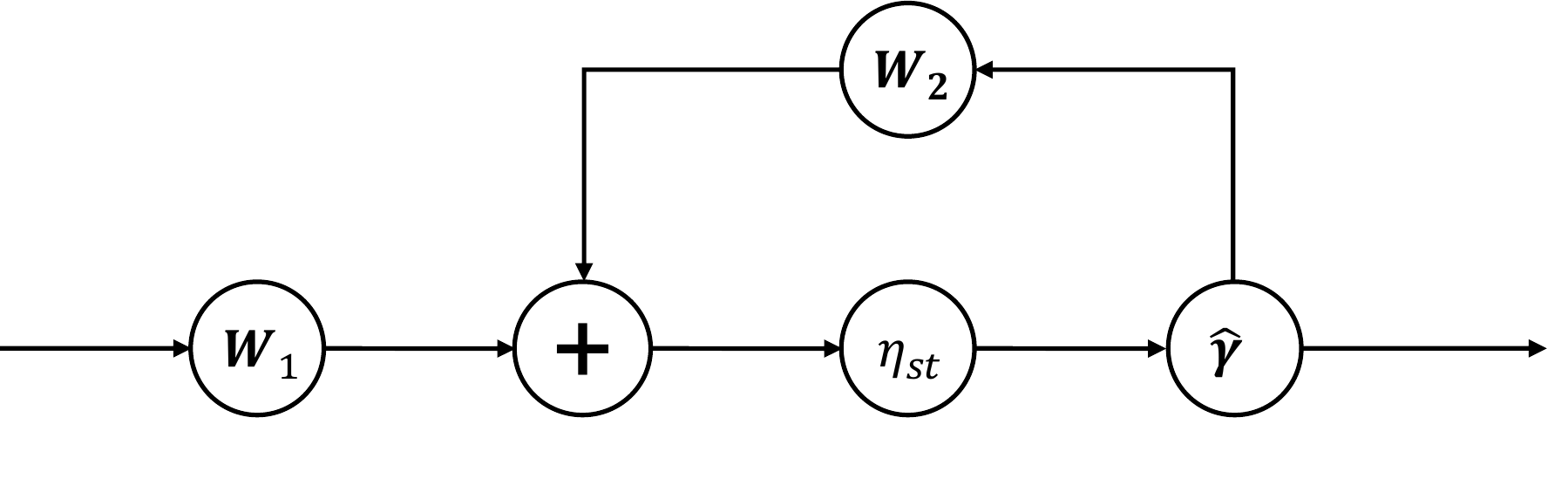}
    \caption{RNN structure of ISTA by viewing an iteration of ISTA as a layer of the RNN.}
    \label{fig:ISTA_RNN}
\end{figure}

If we regard the soft-thresholding function in (\ref{eq:LISTA}) as an activation function,  we find that (\ref{eq:LISTA}) is the basic form of the $i^{th}$ layer of a recurrent neural network (RNN). Therefore, ISTA can be viewed as a RNN illustrated in Fig. \ref{fig:ISTA_RNN}.

\begin{figure*}[h]
    \centering
    \includegraphics[width=0.98\textwidth]{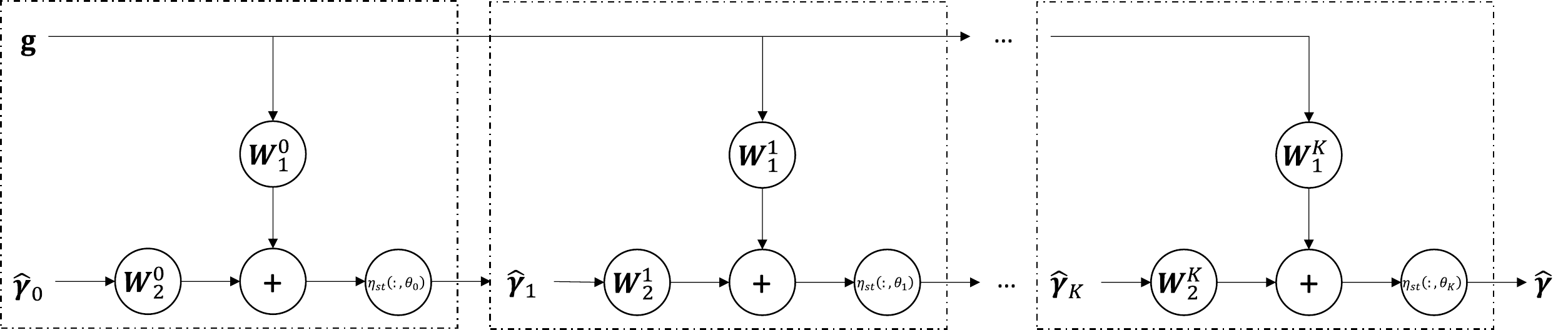}
    \caption{Unfolded LISTA architecture. A \textit{K}-layer LISTA unrolls the RNN and truncates it into \textit{K} iterations, thus leading to a side-connected feedforward neural network.}
    \label{fig:LSITA}
\end{figure*}

Inspired by the connection between ISTA and RNN, a learning based model named Learned ISTA (LISTA) was proposed in \cite{LISTA}. Fig \ref{fig:LSITA} demonstrates the learning architecture of a \textit{K}-layer LISTA, which unrolls the RNN and truncates it into \textit{K} iterations, thus leading to a \textit{K}-layer side-connected feedforward neural network.
The major difference between ISTA and LISTA is that the weight matrices $\mathbf{W}_1^i$, $\mathbf{W}_2^i$ as well as the threshold $\theta_i$ in each layer of LISTA are not pre-determined. Those parameters are learned in the LISTA neural network from training data. 

The loss function of LISTA over the training data $\{ (\mathbf{g}_i, \boldsymbol{\gamma}_i) \}_{i=1}^T$ is the mean square error (MSE) loss described as follows.
\begin{equation}
    \underset{\boldsymbol{\Psi}}{\operatorname{minimize}} \ \mathcal{L}(\boldsymbol{\Psi})=\frac{1}{T} \sum_{i=1}^{T} ||\hat{\boldsymbol{\gamma}}(\boldsymbol{\Psi,\mathbf{g}})-\boldsymbol{\gamma}||_2^2,
    \label{eq:LISTA_train}
\end{equation}
where T denotes the number of samples in the training data and $\boldsymbol{\Psi}=\{\mathbf{W}_1, \mathbf{W}_2, \boldsymbol{\theta}\}$ is the set of free parameters to be learned. Many recent works \cite{LISTA,LISTA2,LISTA3,LISTA4,LISTA5} have demonstrated that LISTA is able to achieve the same estimation accuracy within two to three order-of-magnitude fewer iterations than the original ISTA. Moreover, empirical results show that LISTA has better generalization ability. 

However, to apply LISTA to solve TomoSAR inversion, we need to extend LISTA to complex-valued domain. CV-LISTA shares the same learning architecture as LISTA, except that each neuron in CV-LISTA has two channels, which refer to the real and imaginary part of a complex number, respectively. We applied the following adaptions to the equation (\ref{eq:LISTA})
\begin{equation} 
    \tilde{\boldsymbol{\gamma}}_{i}=\eta_{s t}\left\{\widetilde{\mathbf{W}}_{1}^{i} \tilde{\mathbf{g}}+ \widetilde{\mathbf{W}}_{2}^{i} \tilde{\boldsymbol{\gamma}}_{i-1}, \theta_{i}\right\}
    \label{eq:CV-LISTA}
\end{equation}
where
\begin{align}
\nonumber
\widetilde{\mathbf{W}}_{j}^{i}&=\left[\begin{array}{cc}
\Re(\mathbf{W}_{j}^{i}) & -\Im(\mathbf{W}_{j}^{i}) \\
\Im(\mathbf{W}_{j}^{i}) & \Re(\mathbf{W}_{j}^{i})
\end{array}\right], \\ 
\tilde{\mathbf{g}}&=\left[\begin{array}{l}
\Re(\mathbf{g}) \\
\Im(\mathbf{g})
\end{array}\right], \\ \nonumber
\tilde{\boldsymbol{\gamma}}&=\left[\begin{array}{l}
\Re(\boldsymbol{\hat{\gamma}}) \\
\Im(\boldsymbol{\hat{\gamma}})
\end{array}\right]
\end{align}
with $j = 1,2$ and $\Re(\cdot)$ and $\Im(\cdot)$ denote the real and imaginary operators, respectively.

\subsection{$\boldsymbol{\gamma}$-Net formulation for TomoSAR}


Through our research and experiments, we discovered a few drawbacks of CV-LISTA applying to TomoSAR and proposed several novel improvements in $\boldsymbol{\gamma}$-Net.
The improvements are mainly three-fold. First, the $\mathbf{W}_1$ and $\mathbf{W}_2$ matrices in the aforementioned CV-LISTA are highly correlated. Hence, we introduced weight coupling structure to reduce redundant trainable parameters in $\boldsymbol{\gamma}$-Net.  Second, we tried to use the acceleration technique, support selection, which was originally developed for LASSO to boost the convergence in $\boldsymbol{\gamma}$-Net. Last, we replaced the conventional soft-thresholding function by the piecewise linear function because the conventional soft-thresholding function causes information loss, which leads to large reconstruction error and decreases the convergence rate of the model. We will discuss the improvements in detail in the following paragraphs.

\subsubsection*{\textbf{Weight coupling}}
Instead of training $\boldsymbol{\gamma}$-Net as pure "black-box" networks, we simplify the CV-LISTA and propose $\boldsymbol{\gamma}$-Net by exploiting the dependency among the trainable weights. Details can be found in \cite{LISTA_cpss} that the weights to be learned $\{(\mathbf{W_1^i, W_2^i})\}_{i=0}^{K}$ in each layer asymptotically satisfy the following partial weight coupling structure:
\begin{equation}
    \mathbf{W_{2}^{i}} = \mathbf{I} - \mathbf{W_{1}^{i} R} \ .
    \label{eq:weight_cp}
\end{equation}
 By employing the partial weight coupling structure, we can simplify the $i^{th}$ layer of $\boldsymbol{\gamma}$-Net to:
\begin{align} \label{eq:LISTA_cp}
     \tilde{\boldsymbol{\gamma}}_i = \eta_{st} \{\tilde{\boldsymbol{\gamma}}_{i-1} + \widetilde{\mathbf{W}}^{i} (\tilde{\mathbf{g}} - \widetilde{\mathbf{R}} \tilde{\boldsymbol{\gamma}}_{i-1}), \theta_i\} \\ \nonumber
     \mathrm{with} \ \ \widetilde{\mathbf{R}}= \left[\begin{array}{cc}
    \Re(\mathbf{R}) & -\Im(\mathbf{R}) \\
    \Im(\mathbf{R}) & \Re(\mathbf{R})
\end{array}\right]
\end{align}
where $(\Re(\mathbf{W^{i}}), \Im(\mathbf{W^{i}}), \theta^i)$ are the parameters to be learned in the $i^{th}$ layer, and the trainable weight $\mathbf{W^i}$ is initialized using the system measurement matrix $\mathbf{R}$ with $\mathbf{W^i} = \beta \mathbf{R}^H$. The coupled structure contributes to eliminating the number of free parameters to be trained, thus accelerating the training procedure significantly. Theoretically speaking, (\ref{eq:weight_cp}) can only be satisfied for deep layers. However, extensive simulations in \cite{LISTA_cpss} demonstrate that the application of the partial weight coupling structure to every layer will not degrade the theoretical and experimental performance. 

\subsubsection*{\textbf{Support selection}}
In addition to the application of the weight coupling structure, we introduce a special thresholding scheme in $\boldsymbol{\gamma}$-Net, called \textit{support selection}, which is inspired by ``kicking" in linearized Bregman iteration \cite{osher2011fast}. Meaning, we will select a certain percentage of entries with largest magnitude at each layer of $\boldsymbol{\gamma}$-Net before the shrinkage step. Hereafter, the selected part will be trusted as "true support" and directly fed into the following layer, bypassing the shrinkage step. The remaining entries will go through the shrinkage step as usual. Assuming that $\rho^i$ percentage of entries are trusted in the $i^{th}$ layer, the support selection can be formally defined as:
\begin{equation}
    {\eta_{ss}}_{\theta_i}^{\rho^i}(\tilde{\boldsymbol{\gamma}}_i)=\left \{
    \begin{array}{lr}
         \tilde{\boldsymbol{\gamma}}_i  & i \in \mathcal{S}^{\rho^i}(\tilde{\boldsymbol{\gamma}}) \\
         \eta_{st}(\tilde{\boldsymbol{\gamma}}_i, \theta_i)  & i \notin \mathcal{S}^{\rho^i}(\tilde{\boldsymbol{\gamma}})
    \end{array}
    \right. ,
    \label{eq:ss}
\end{equation}
where $\mathcal{S}^{\rho^i}(\boldsymbol{\gamma})$ contains the entries with the $\rho^i$ largest magnitudes. It is worth mentioning that the percentage $\rho^i$ is a hyperparameter that requires manual tuning. When we apply the support selection to $\boldsymbol{\gamma}$-Net, then (\ref{eq:LISTA_cp}) is modified as:
\begin{equation}
     \tilde{\boldsymbol{\gamma}}_i = {\eta_{ss}}_{\theta_i}^{\rho^i} \{\tilde{\boldsymbol{\gamma}}_{i-1} + \widetilde{\mathbf{W}}^{i} (\tilde{\mathbf{g}} - \widetilde{\mathbf{R}} \tilde{\boldsymbol{\gamma}}_{i-1}), \theta_i\}
     \label{eq:lista_cpss}
\end{equation}
Simulation experiments in \cite{LISTA_cpss} support that introducing the support selection on the one hand improves the convergence rate both theoretically and empirically, on the other hand, it contributes to reducing the recovery error, thus improving the estimation accuracy.

\subsubsection*{\textbf{Piecewise linear thresholding function}}
The conventional soft-thresholding function, simply prunes elements with small magnitude to zero, which is very likely to result in information loss. In order to maintain useful information as much as possible and execute the shrinkage step in the meanwhile, we replace the soft-thresholding function by the piecewise linear function $\eta_{p w l}(\tilde{\boldsymbol{\gamma}}, \boldsymbol{\theta})$ in $\boldsymbol{\gamma}$-Net, which is defined as:
\begin{equation}
\eta_{p w l}(\hat{\boldsymbol{\gamma}})=
   \left\{\begin{array}{ll}
\theta_{3} \hat{\boldsymbol{\gamma}}, & |\hat{\boldsymbol{\gamma}}| \leq \theta_{1} \\ 
\\
e^{j \cdot \angle \hat{\boldsymbol{\gamma}}} [\theta_{4}(|\hat{\boldsymbol{\gamma}}|-\theta_{1})+ \\ 
\qquad \quad \theta_{3} \theta_{1}], &
\theta_{1}<|\hat{\boldsymbol{\gamma}}| \leq \theta_{2}\\
\\
e^{j \cdot \angle \hat{\boldsymbol{\gamma}}} [\theta_{5}(|\hat{\boldsymbol{\gamma}}|-\theta_{2}) + \\ \theta_{4}\left(\theta_{2}-\theta_{1}\right)+\theta_{3} \theta_{1}], & |\hat{\boldsymbol{\gamma}}|>\theta_{2} 

\end{array}\right. .
\label{eq:piecewise}
\end{equation}
To clarify, the symbol $j$ in the equation (\ref{eq:piecewise}) refers to the imaginary number.
\begin{figure}[h]
    \centering
    \includegraphics[width=0.49\textwidth]{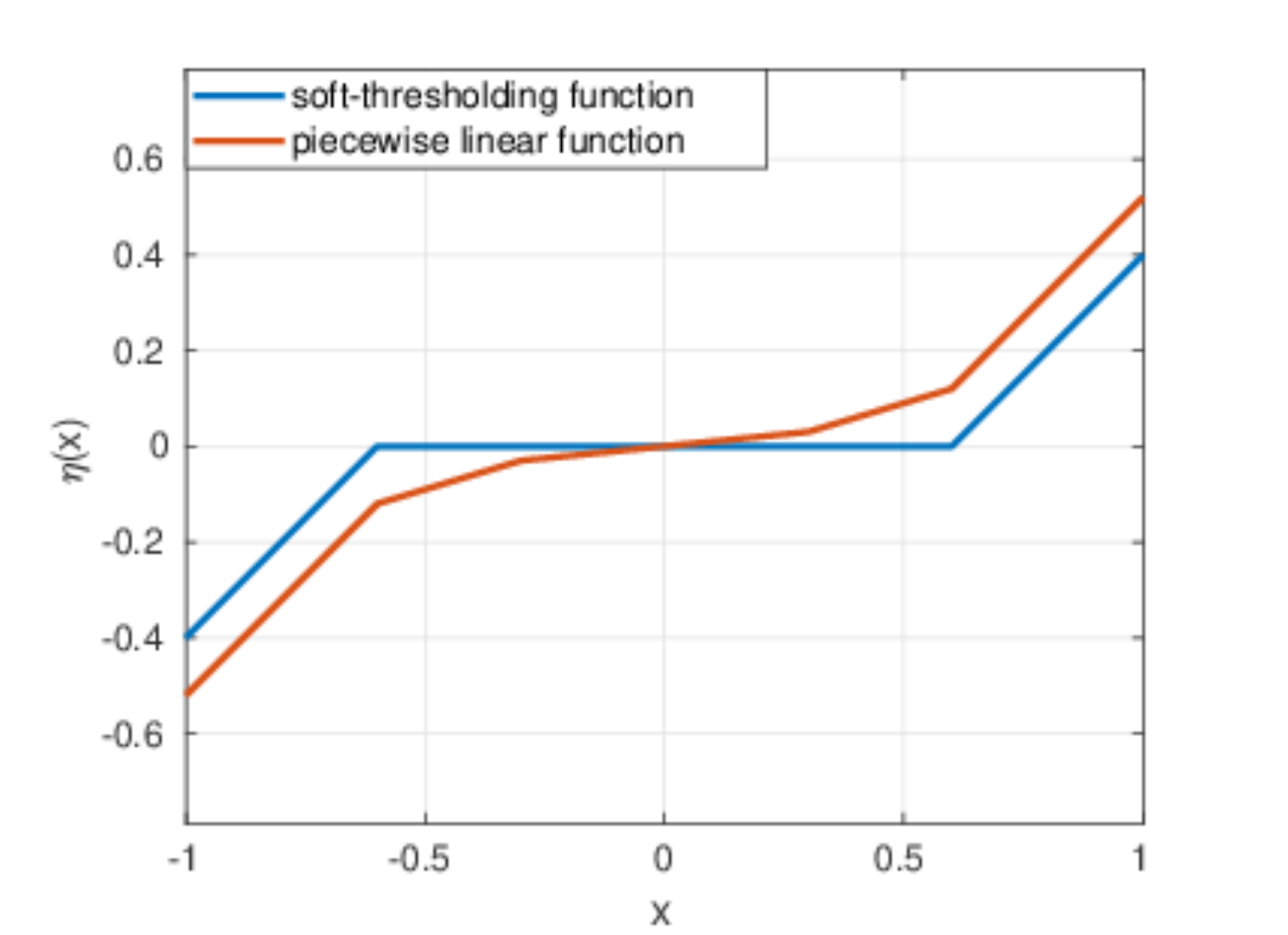}
    \caption{Comparison between the piecewise linear function and soft-thresholding function. Instead of pruning the elements with small magnitude, the piecewise linear function just further minifies them, thus possibly avoiding the information loss.}
    \label{fig:pwl_st}
\end{figure}

Fig. \ref{fig:pwl_st} compares both functions. As can be seen, instead of pruning elements with small magnitude, the piecewise linear function only down scale them. Hence, it mitigates the information loss. However, it leads to the consequence that the final output of $\boldsymbol{\gamma}$-Net being not strictly sparse. Most elements of the final output are not driven to zero strictly but to some extremely small values. Therefore, an additional post-processing step for cleaning the elements with extremely small magnitude is necessary when we employ the piecewise linear function. 

\begin{figure}[h]
    \centering
    \includegraphics[width=0.49\textwidth]{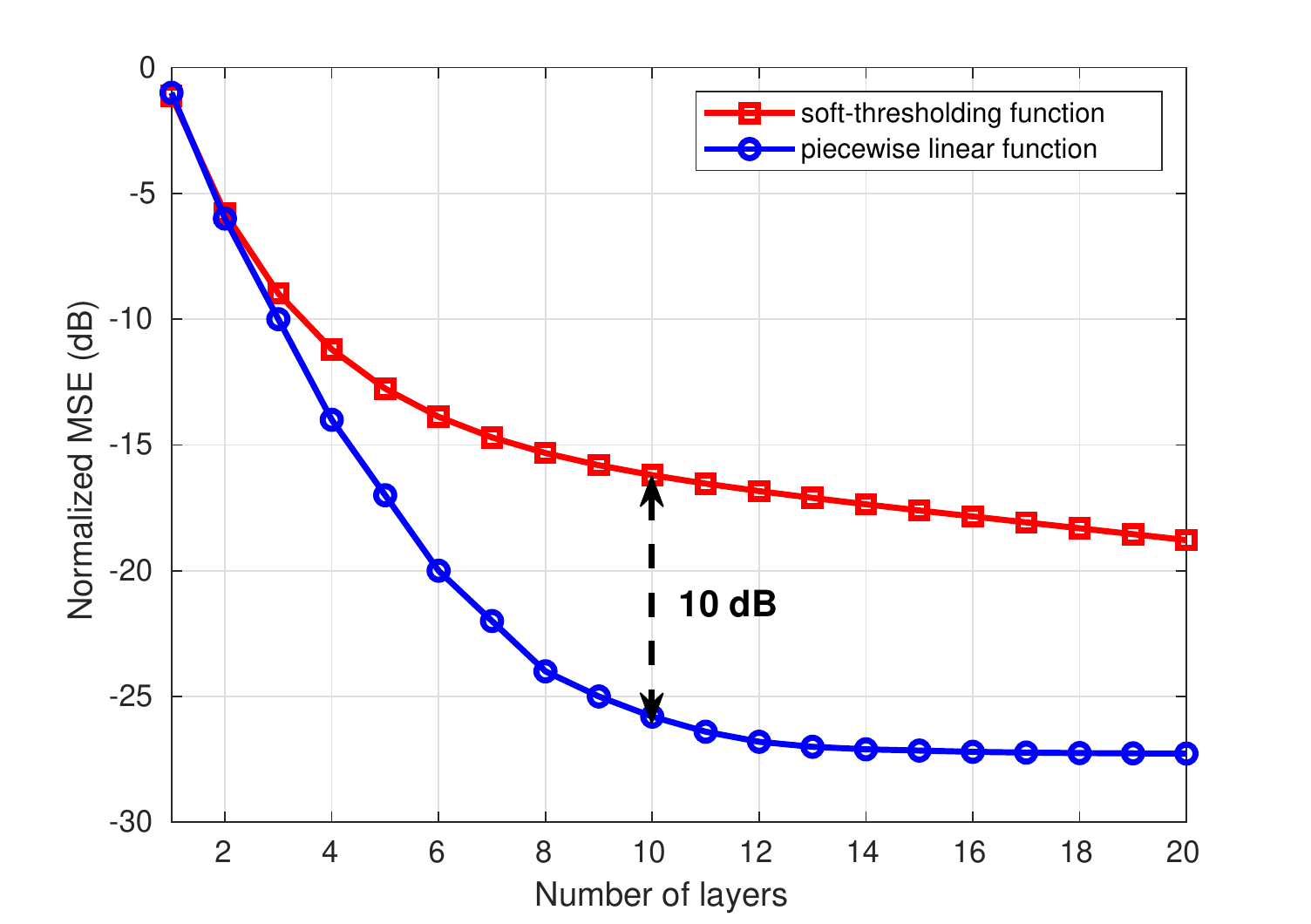}
    \caption{Performance of $\boldsymbol{\gamma}$-Net using different shrinkage function. The piecewise linear function conduces to faster convergence and improves the estimation accuracy.}
    \label{fig:pwl_st_db}
\end{figure}
Fig. \ref{fig:pwl_st_db} compares $\boldsymbol{\gamma}$-Net performance in term of the normalized mean square error (NMSE) under the two shrinkage functions. To clarify, the performance of $\boldsymbol{\gamma}$-Net with different shrinkage functions was verified on a set of noise free data so that the results reflect the ideal performance. The NMSE is defined as:
\begin{equation}
    \mathrm{N M S E}=\frac{1}{T} \sum \frac{\|\hat{\boldsymbol{\gamma}}-\boldsymbol{\gamma}\|_{2}^{2}}{\|\boldsymbol{\gamma}\|_{2}^{2}}
\end{equation}

From this figure, it can be seen that $\boldsymbol{\gamma}$-Net with the piecewise linear function achieves lower NMSE at the same number of layers. In another word, the piecewise linear function improves the estimation accuracy of $\boldsymbol{\gamma}$-Net or the convergence rate. Specifically, $\boldsymbol{\gamma}$-Net with the piecewise linear function requires only about 12 layers to achieve convergence. However, it is obvious that much more layers are required when the conventional soft-thresholding function is employed.

\subsection{Algorithm summary}
In order to achieve super-resolution ability and high elevation estimation accuracy, it is usually required to sample the elevation range much denser than the elevation resolution unit, thus rendering the steering matrix $\mathbf{R}$ severely over-complete, reducing its restricted isometric property (RIP) and increasing its coherence. The violation of RIP and incoherence introduces outliers to estimates of the reflectivity profile $\hat{\boldsymbol{\gamma}}$ \cite{dlr81574}. Additionally, outliers will be caused by the noise interference as well. Hence, we need further perform model order selection \cite{Zhu2010Very} to suppress the undesired outliers and estimate the number and location of scatterers precisely, which are typical steps in TomoSAR. The proposed super-resolving TomoSAR inversion algorithm is a combination of $\boldsymbol{\gamma}$-Net and model order selection, and re-estimation. The basic workflow of the proposed algorithm is show in Algorithm \ref{table:algorithm}. The model order selection is conducted based on Bayesian Information Criterion (BIC) \cite{BIC}. 

\begin{algorithm}[h]
\caption{Summary of the proposed algorithm}
\label{table:algorithm}
\begin{algorithmic}
    \STATE \textbf{Simulate training data} \\
         \STATE \qquad  \textbf{Sampling} the elevation extent \\
         \STATE \qquad  \textbf{Generate} steering matrix $\mathbf{R}$ with \\
         \STATE \qquad $R_{n l}=\exp \left(-j 2 \pi \xi_{n} s_{l}\right)$, where $\xi_{n}=-2 b_{n} /(\lambda r)$\\
         \STATE \qquad \textbf{Simulate} reflectivity profile $\boldsymbol{\gamma}$ \\
         \STATE \qquad \textbf{Simulate} SAR measurements with $\mathbf{g = R} \boldsymbol{\gamma}+ \boldsymbol{\varepsilon}$  \\
         \STATE \qquad \textbf{Finish} the generation of training data $\{(\mathbf{g}_i, \boldsymbol{\gamma}_i)\}_{i=1}^{T}$ \\
     \STATE \textbf{Training of $\boldsymbol{\gamma}$-Net} \\
     \STATE \qquad \textbf{Over given training samples} $\{ (\mathbf{g_i, \boldsymbol{\gamma}_i}) \}_{i=1}^{T}$ \\
     \STATE \qquad $\underset{\boldsymbol{\Psi}}{\operatorname{\textbf{minimize}}} \ \mathcal{L}(\boldsymbol{\Psi})=\frac{1}{T} \sum_{i=1}^{T} ||\hat{\boldsymbol{\gamma}}(\boldsymbol{\Psi,\mathbf{g}})-\boldsymbol{\gamma}||_2^2$  \\
     \STATE \qquad where $\boldsymbol{\Psi}=[\Re(\mathbf{W}), \Im(\mathbf{W}), \boldsymbol{\theta}]$
     \FOR {each pixel in the image:} 
         \STATE \qquad \textbf{Preliminary estimate} via $\boldsymbol{\gamma}$-Net: \\ \STATE \qquad $\boldsymbol{\gamma}$ = $\boldsymbol{\gamma}$-Net($\mathbf{g}$) \\
         \STATE \qquad \textbf{Model order selection} to remove outliers:
         \STATE \qquad $\hat{P}=\underset{P}{\operatorname{argmin}}\left\{\sigma_{\varepsilon}^{-2}|| \mathbf{g}-\mathbf{R} \hat{\gamma} \|_{2}^{2}+1.5 P \ln N\right\}$ \\
         \STATE \qquad \textbf{Determine the number of scatterers}
         \STATE \qquad \textbf{Final estimation} of their elevation
         \ENDFOR
 \end{algorithmic}
\end{algorithm}



\section{Performance Evaluation}
\subsection{Data simulation and training}
\subsubsection*{\textbf{Simulation setup}}

We simulate the data using a setting similar to \cite{Zhu2012Super-Resolution}. Specifically, SAR measurements with 25 spatial baselines are simulated. The spatial baselines are regularly distributed in the range from -135m to 135m, thus leading to a Rayleigh resolution of around 42m. We simulated ca. 4 million training samples, half of which are single scatterer and the others are overlaid double scatterers. The simulation details of single and double scatterers are listed below. We randomized many parameters in the simulation, in order to make the simulation more realistic. 

\begin{itemize}[leftmargin=*]
    \item Single scatterers: for a single scatterer, the scatterering coefficient is a complex number $\gamma = A \cdot \exp{(j\phi)}$, with the amplitude being deterministic and the scattering phase following an uniform distribution, i.e. $\phi \sim U(0,2\pi)$. In order to randomize the amplitude in the simulation, we simulate it with a uniform distribution as well, i.e. $A \sim U(1,4)$, although a real SAR amplitude image shows more Rayleigh or Gamma distribution. The elevation of the simulated scatterer is regularly distributed in the range from 0m to 200m with 1m sampling. Once the location of the scatterer is determined, the echo signal is simulated at 11 different levels of SNR, which is regularly distributed between [0dB, 10dB].\\
    \item Double scatterers: For double scatterers, the generation of the two single scatterers is identical to the previous step, i.e. for each individual scatterer, the phase follows an uniform distribution $\phi \sim U(0,2\pi)$ and the amplitude follows an uniform distribution $A \sim U(1,4)$, respectively. As a result, different amplitude ratio and phase difference of the simulated double scatterers can be covered. We also vary the elevation distance between the two single scatterers. The elevation distance varies from 0.1 until 1.2 Rayleigh resolution, with a regular sampling of 0.1 Rayleigh resolution. The elevation of the first scatterer follows an uniform distribution in the range of 0m to 200m. In order to avoid the off-grid bias, we assume that all scatterers locate on-grid with 1m sampling. 

\end{itemize}

\subsubsection*{\textbf{Training}}
\begin{figure}[h]
    \centering
    \includegraphics[width=0.49\textwidth]{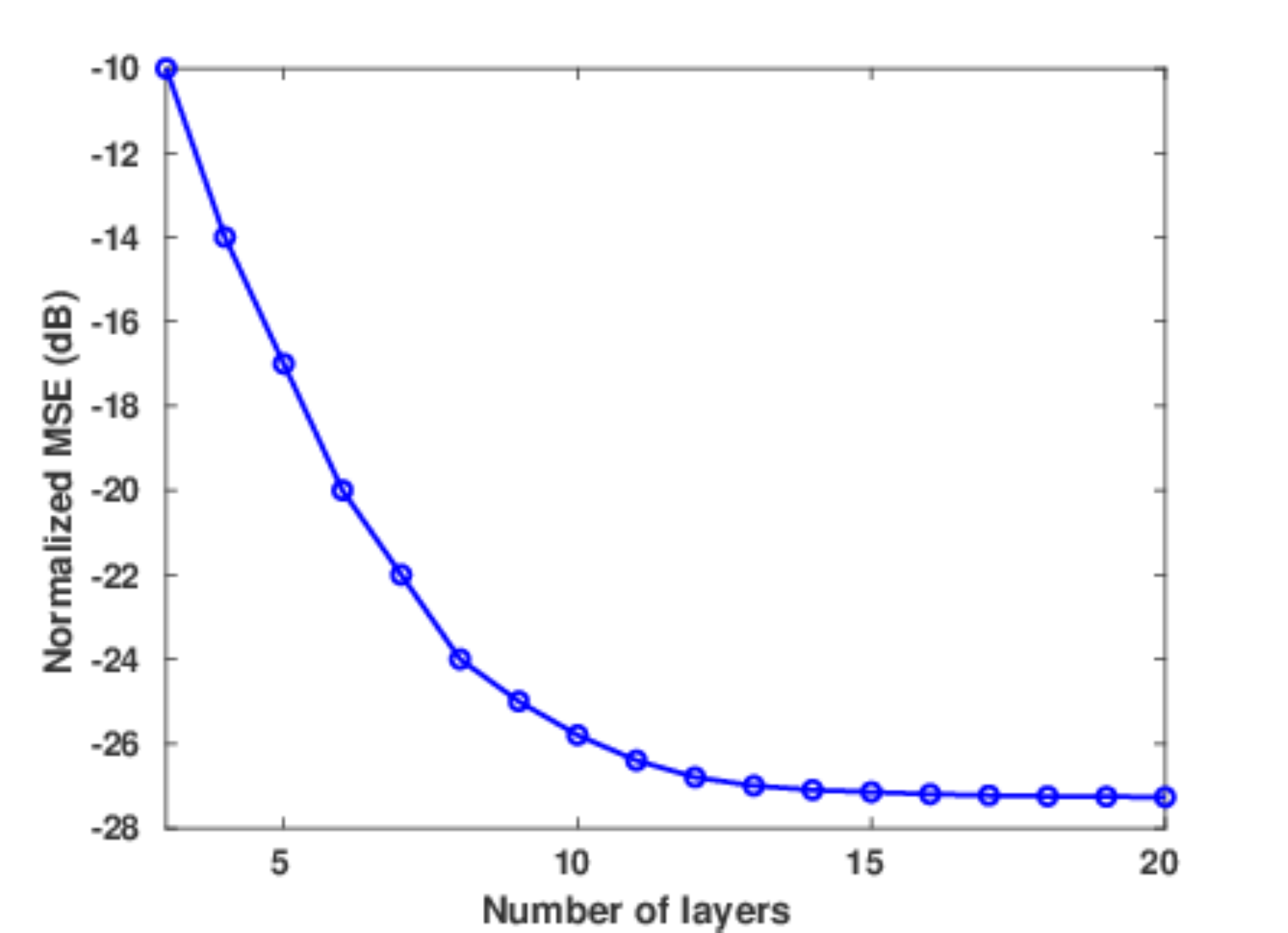}
    \caption{$\boldsymbol{\gamma}$-Net performance w.r.t the number of layers. After 12 layers, the performance improvement of $\boldsymbol{\gamma}$-Net is marginal with the increase number of layers. Instead, the increase of layers leads to heavier computational burden.}
    \label{fig:loss_layers}
\end{figure}

The training was carried out using Pytorch \cite{pytorch} and the Adam optimizer \cite{adam}. The learning rate was initialized at 0.0005 and adjusted adaptively during the training. In the training procedure, we gradually increase the number of the layers from 3 to 20 in order to determine an optimal network structure. We validate the performance of $\boldsymbol{\gamma}$-Net with different number of layers on a validation dataset. The validation dataset contains 0.2 million noise-free samples simulated using the same settings mentioned in the simulation setup, so that we can compare the theoretical performance of $\boldsymbol{\gamma}$-Net with different number of layers. Fig. \ref{fig:loss_layers} illustrates the performance of $\boldsymbol{\gamma}$-Net w.r.t the number of its layers. Closer inspection of Fig. \ref{fig:loss_layers} shows that the NMSE firstly decreases rapidly and then starts to converge to a minimum at around twelve layers. Simultaneously, the increase of the number of layers leads to heavier computation cost. Therefore, $\boldsymbol{\gamma}$-Net employed in this paper contains just twelve layers. On the one hand, $\boldsymbol{\gamma}$-Net with twelve layers is able to guarantee the estimation accuracy; on the other hand, it maintains the computational efficiency.

\subsection{\textbf{Single scatterer analysis}}
\begin{figure*}[h]
    \centering
    \begin{minipage}[t]{0.48\linewidth}
    \includegraphics[width=0.98\textwidth]{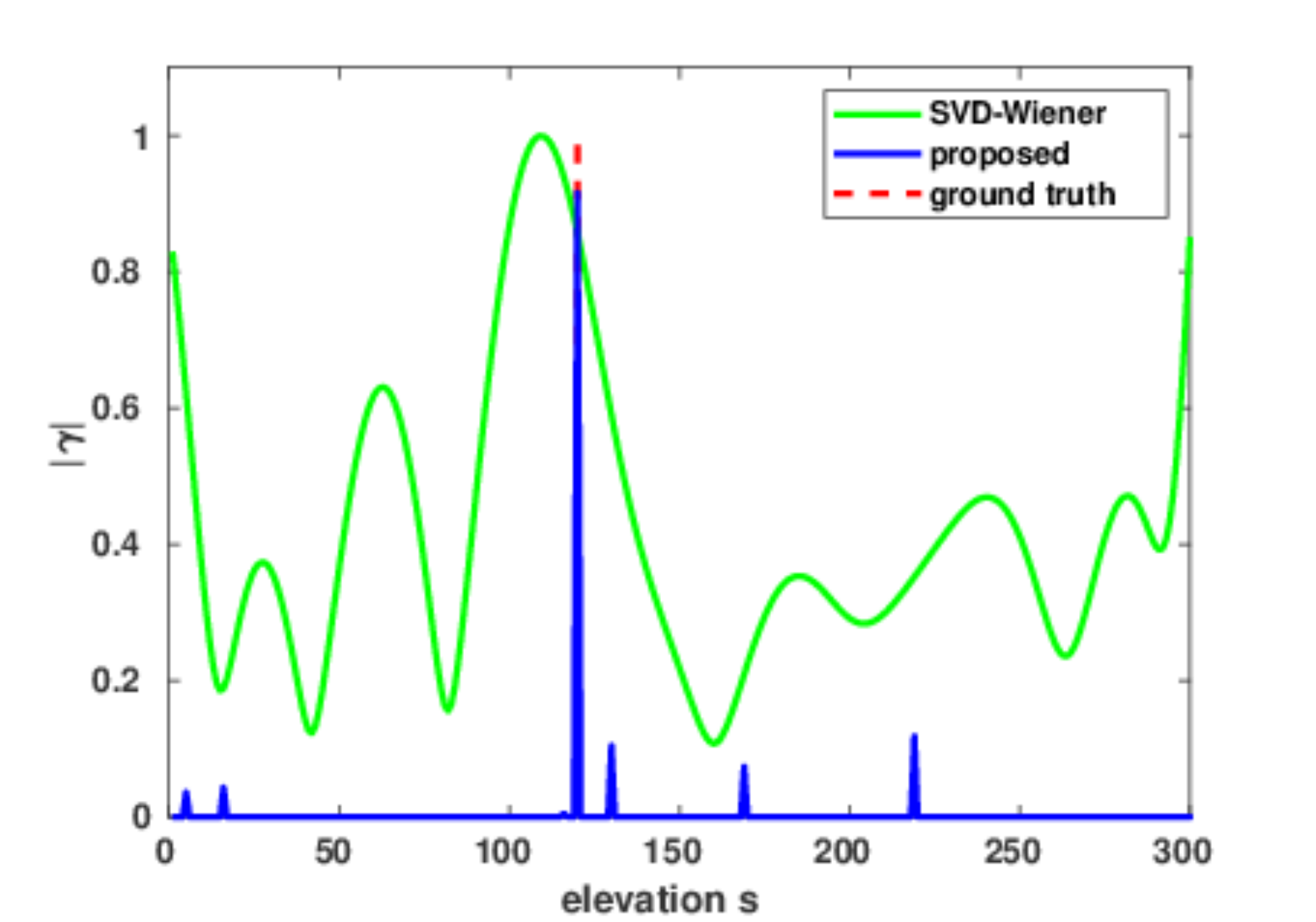}
    \caption*{(a)}
    \end{minipage}
    \begin{minipage}[t]{0.48\linewidth}
    \includegraphics[width=0.98\textwidth]{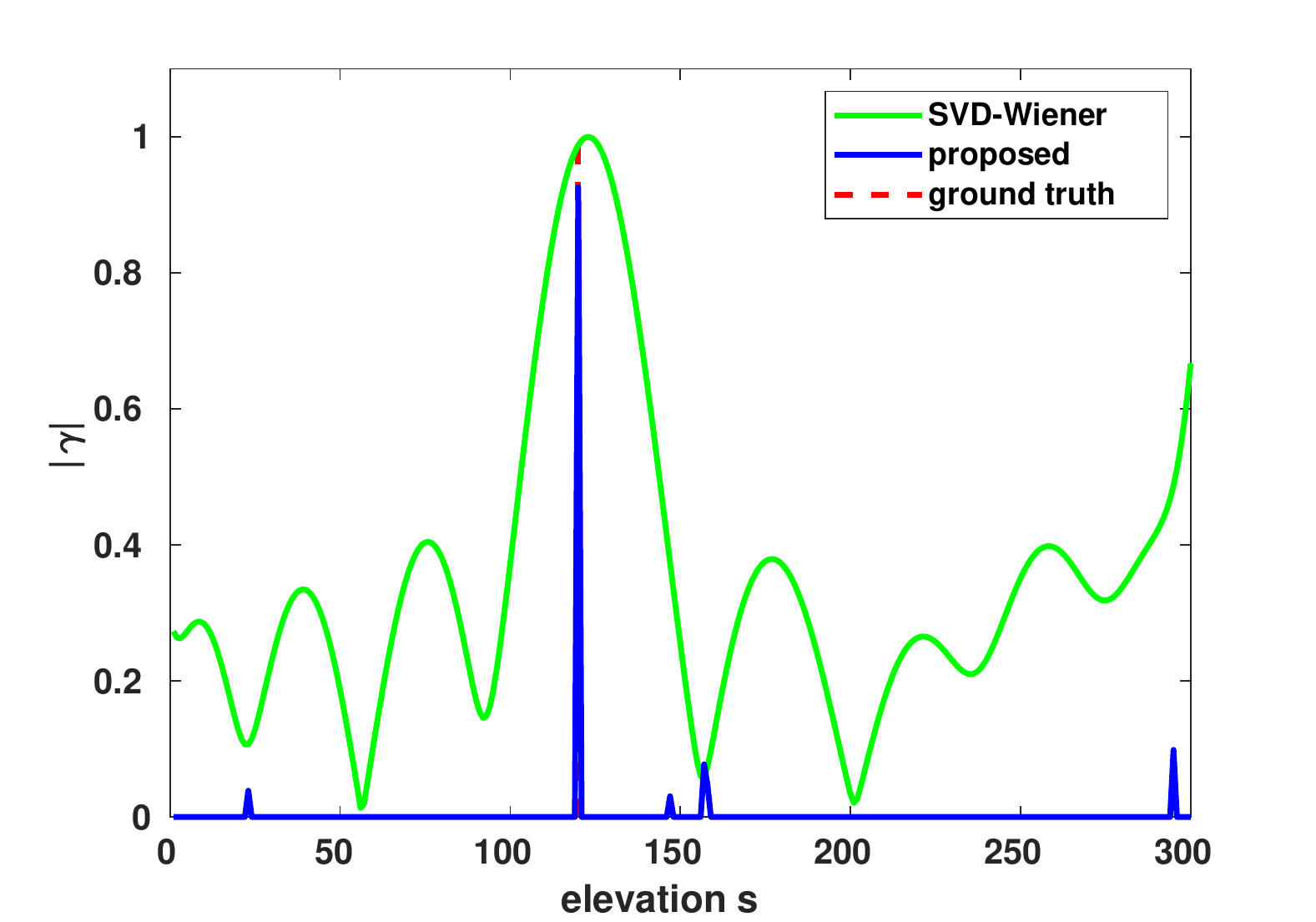}
    \caption*{(b)}
    \end{minipage}
    \begin{minipage}[t]{0.48\linewidth}
    \includegraphics[width=0.98\textwidth]{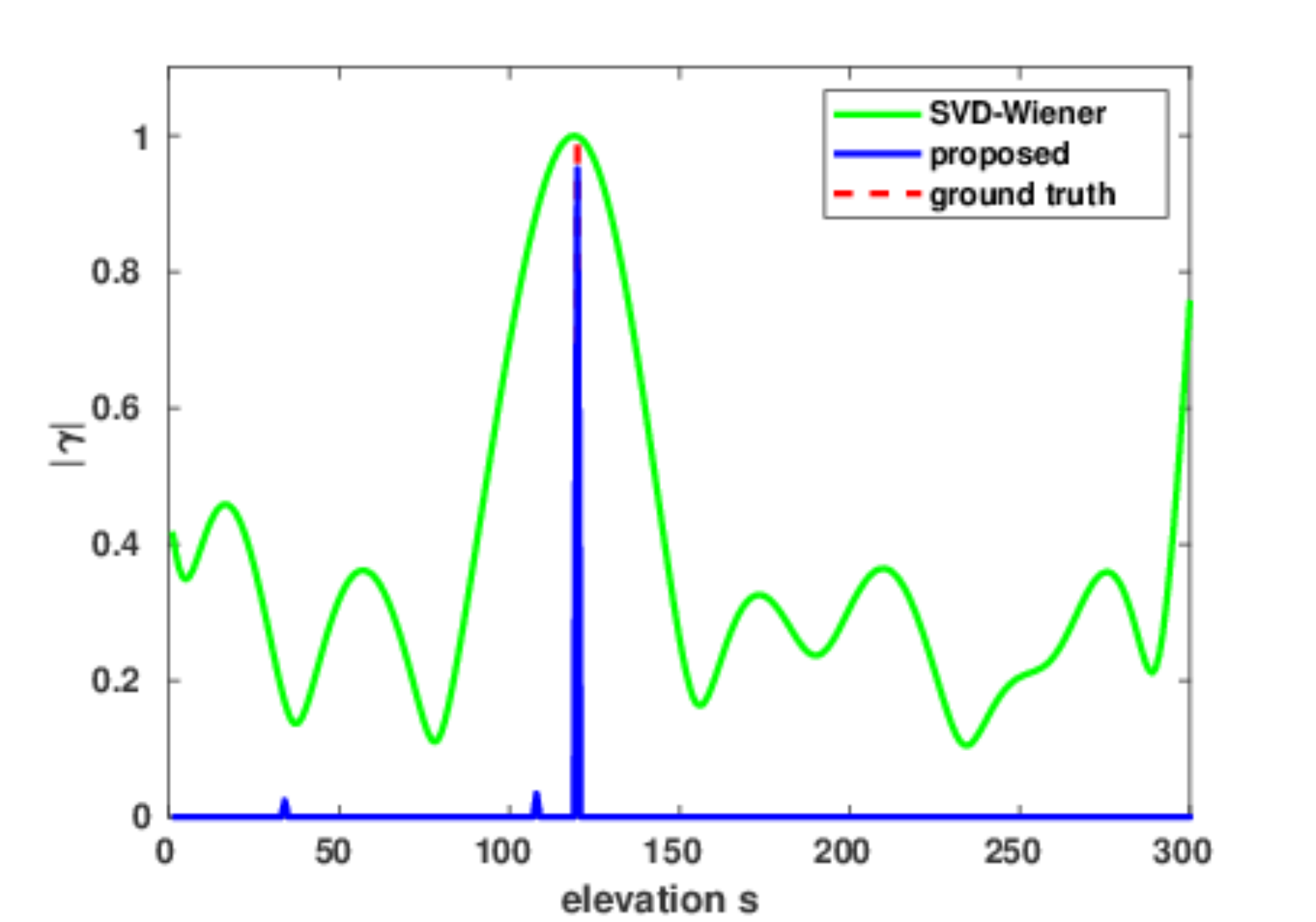}
    \caption*{(c)}
    \end{minipage}
    \begin{minipage}[t]{0.48\linewidth}
    \includegraphics[width=0.98\textwidth]{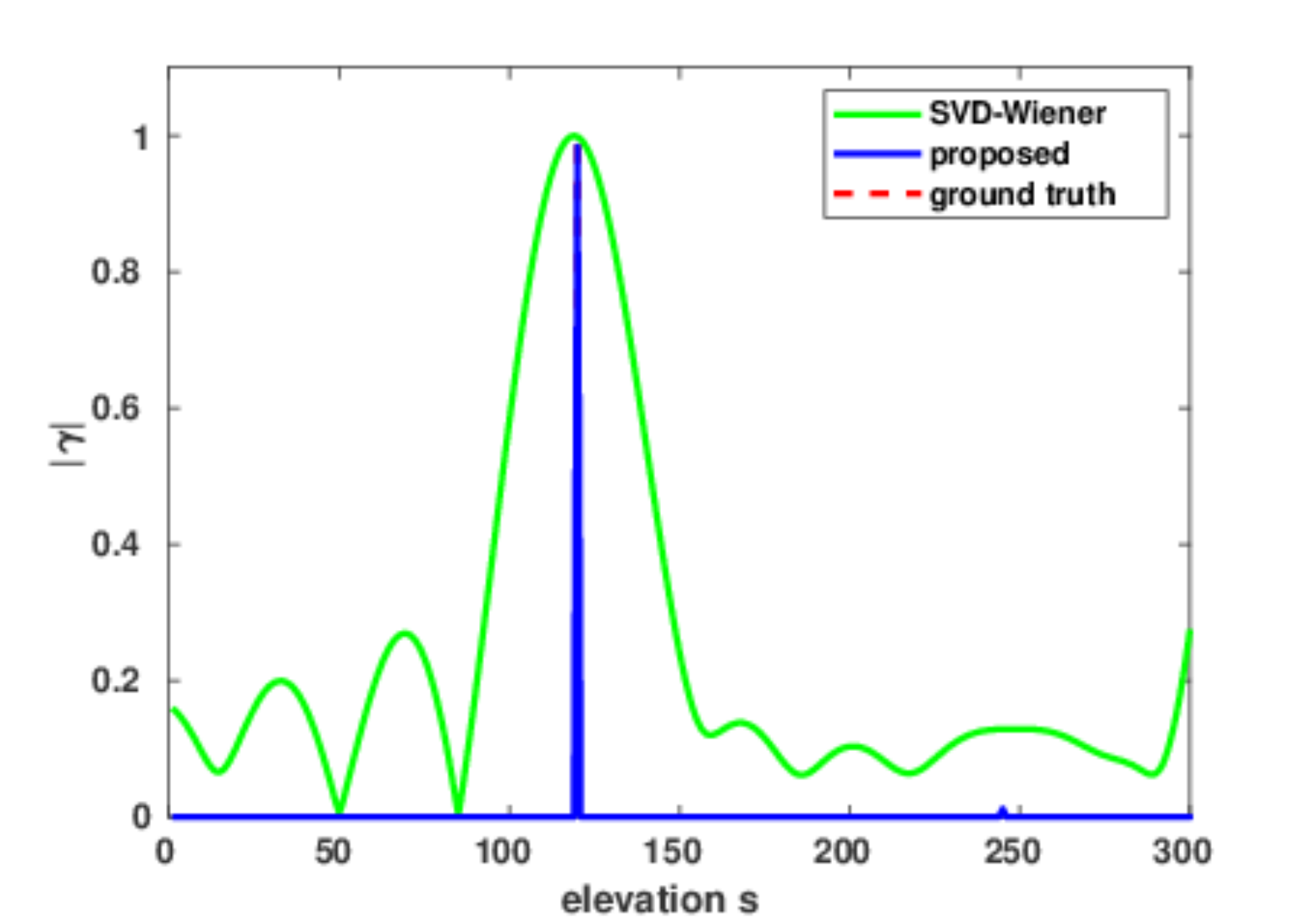}
    \caption*{(d)}
    \end{minipage}
    \caption{Estimated reflectivity profile of simulated data with single scatterer at different SNR. (a) $SNR=0$dB, (b)$SNR=3$dB, (c) $SNR=6$dB, (d)$SNR=10$dB.}
    \label{fig:comp_sing}
\end{figure*}

In addition to the simulated training data, we simulated four sets of testing data for the single scatterer analysis with SNR=$\{0,3,6,10\}$dB. Each set is composed of 0.2 million samples. We use the proposed algorithm to detect the single scatterer and estimate the corresponding elevation. Fig. \ref{fig:comp_sing} demonstrates the estimated reflectivity profile using the trained $\boldsymbol{\gamma}$-Net and SVD-Wiener \cite{Zhu2010Very} (a conventional non-superresolving algorithm). As we can see, although both of $\boldsymbol{\gamma}$-Net and SVD-Wiener are able to detect the position of the single scatterer, $\boldsymbol{\gamma}$-Net reconstructs spectral lines instead of sinc-like point response function, thus mitigating the sidelobe problem. Moreover, from Fig. \ref{fig:comp_sing}(a)-(d), we can see that the outliers caused by noise interference exist in the reflectivity profile estimate of $\boldsymbol{\gamma}$-Net. Therefore, further model order selection step is required 

Table \ref{table:sing_sca} provides the results after model order selection. The CRLB, the estimates mean $(\mu)$ and standard deviation $(\sigma)$ in Table \ref{table:sing_sca} are normalized to the Rayleigh resolution. Since the goal of TomoSAR is to have a good elevation estimate, and also a high detection rate, we define the term \textit{\textbf{effective detection rate}}. An effective detection of single scatterer should satisfy the following two conditions: (1) only one scatterer is detected; (2) the estimated elevation should not exceed $\pm 3$ times CRLB w.r.t the ground truth. 
It is apparent from this table that the proposed algorithm is able to detect almost all single scatterer at different SNRs. Further statistics on mean value $\mu$ and standard deviation $\sigma$ of the estimation error indicate high estimation accuracy of the proposed algorithm with the bias approaching zero and the standard deviation approaching the CRLB.

\begin{table*}[h]
\centering
\begin{tabular}{p{0.15\textwidth} p{0.17\textwidth} p{0.17\textwidth} p{0.17\textwidth} p{0.17\textwidth}}
\toprule
SNR (dB)  & effective detection rate  &  CRLB (normalized) & $\sigma$ (normalized) &  $\mu$ (normalized)    \\
\midrule
0 & $94.19 \%$ & $7 \times 10^{-2}$ & $9 \times 10^{-2}$ & $9 \times 10^{-3}$  \\
 3 & $96.34 \%$ & $5 \times 10^{-2}$ & $6 \times 10^{-2}$ & $5 \times 10^{-3}$  \\
 6 & $98.81 \%$ & $3 \times 10^{-2}$ & $3 \times 10^{-2}$ & $2 \times 10^{-3}$  \\
 10 & $99.79 \%$ & $2 \times 10^{-2}$ & $2 \times 10^{-2}$ & $6 \times 10^{-4}$  \\
\bottomrule
\end{tabular}
\caption{Statistics of the estimate of single scatterer using the proposed algorithm. $\mu$ and $\sigma$ denote the sample mean and the corresponding standard deviation,respectively. The proposed algorithm is able to detect the single scatterer in nearly all cases with the standard deviation approaching the CRLB and bias approaching zero.} 
\label{table:sing_sca}
\end{table*}

\subsection{\textbf{Double scatterers analysis}}

\begin{figure*}[h!]
    \centering
    \begin{minipage}[t]{0.32\linewidth}
    \includegraphics[width=0.98\textwidth]{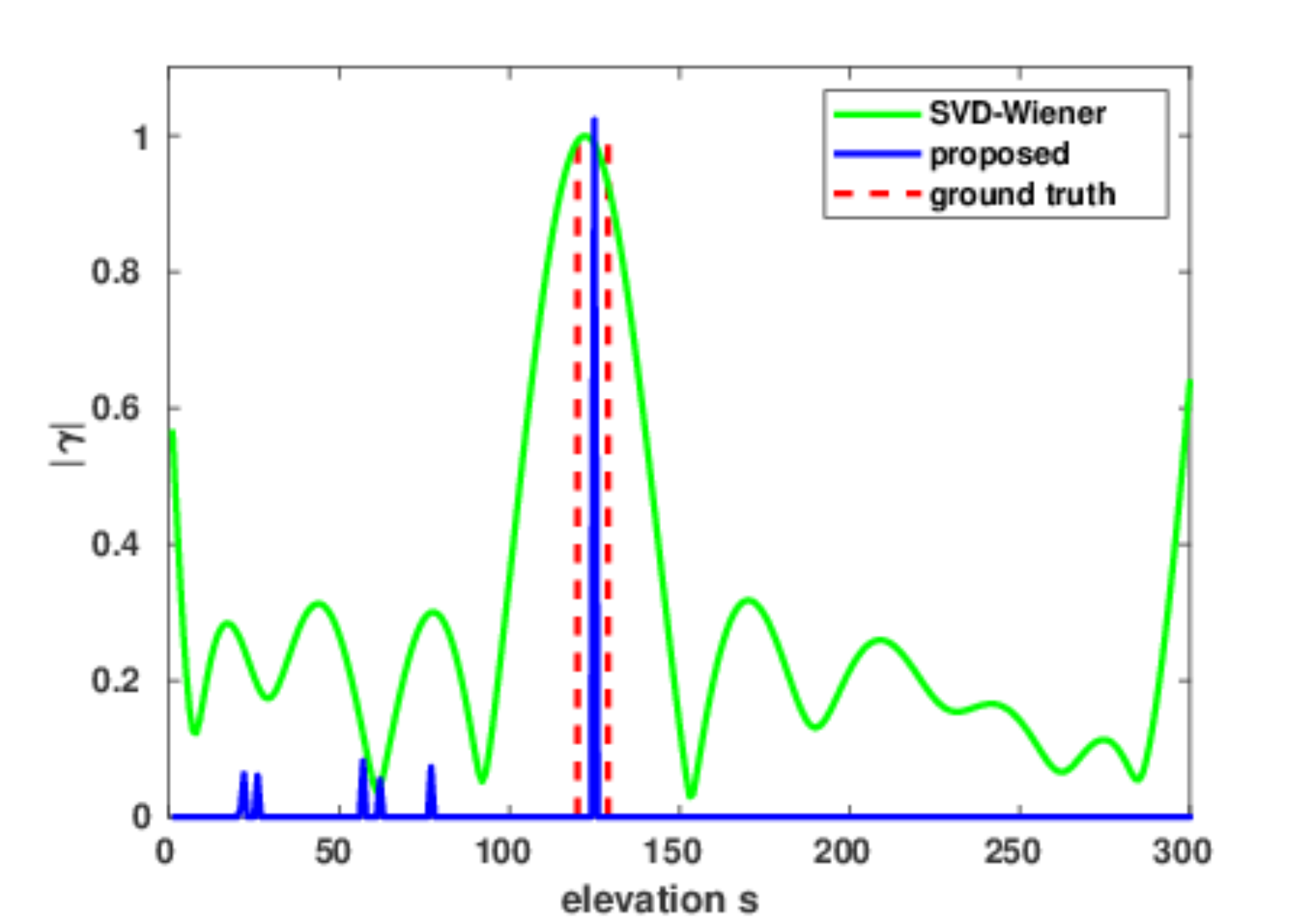}
    \caption*{(a)}
    \end{minipage}
    \begin{minipage}[t]{0.32\linewidth}
    \includegraphics[width=0.98\textwidth]{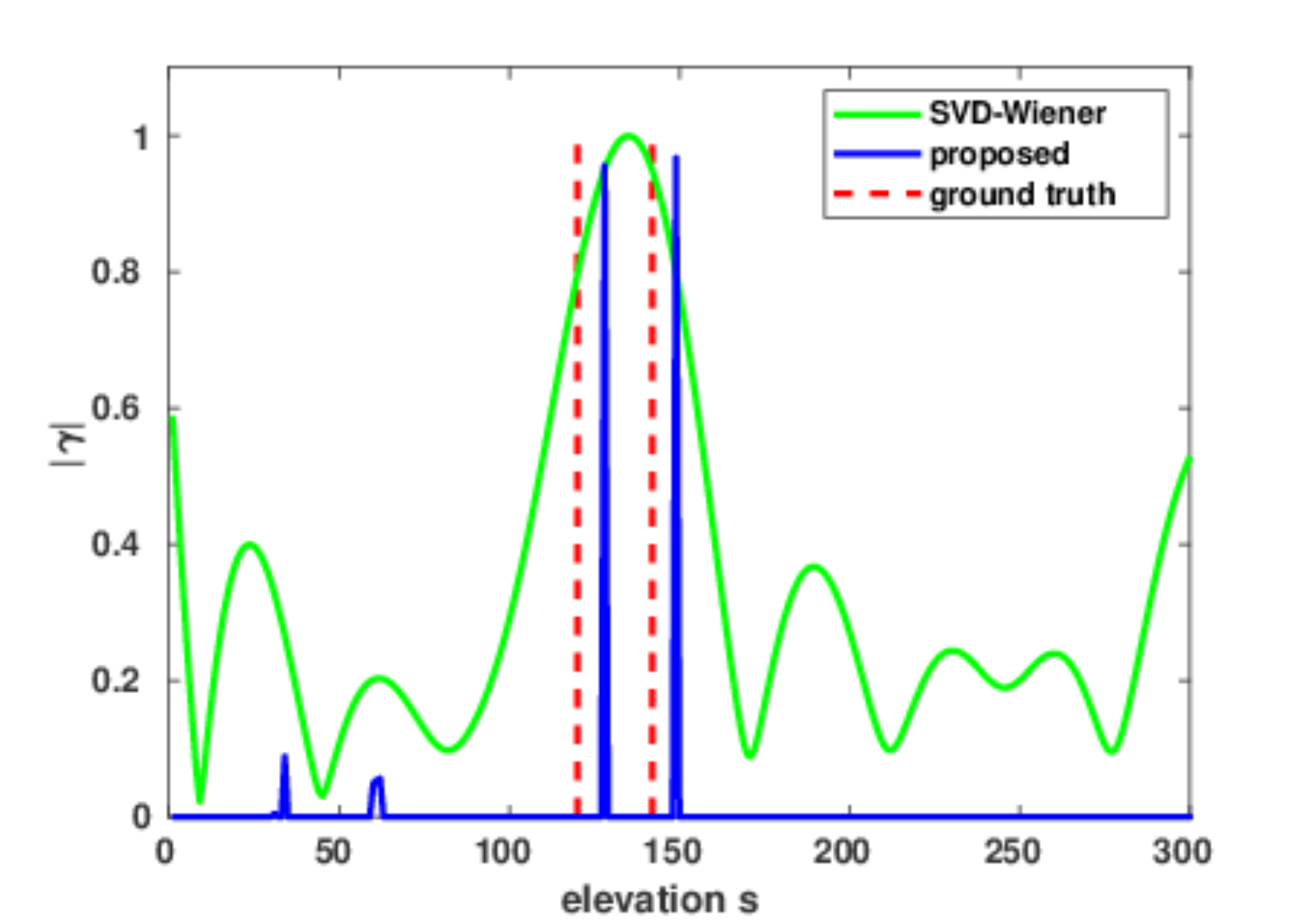}
    \caption*{(b)}
    \end{minipage}
    \begin{minipage}[t]{0.32\linewidth}
    \includegraphics[width=0.98\textwidth]{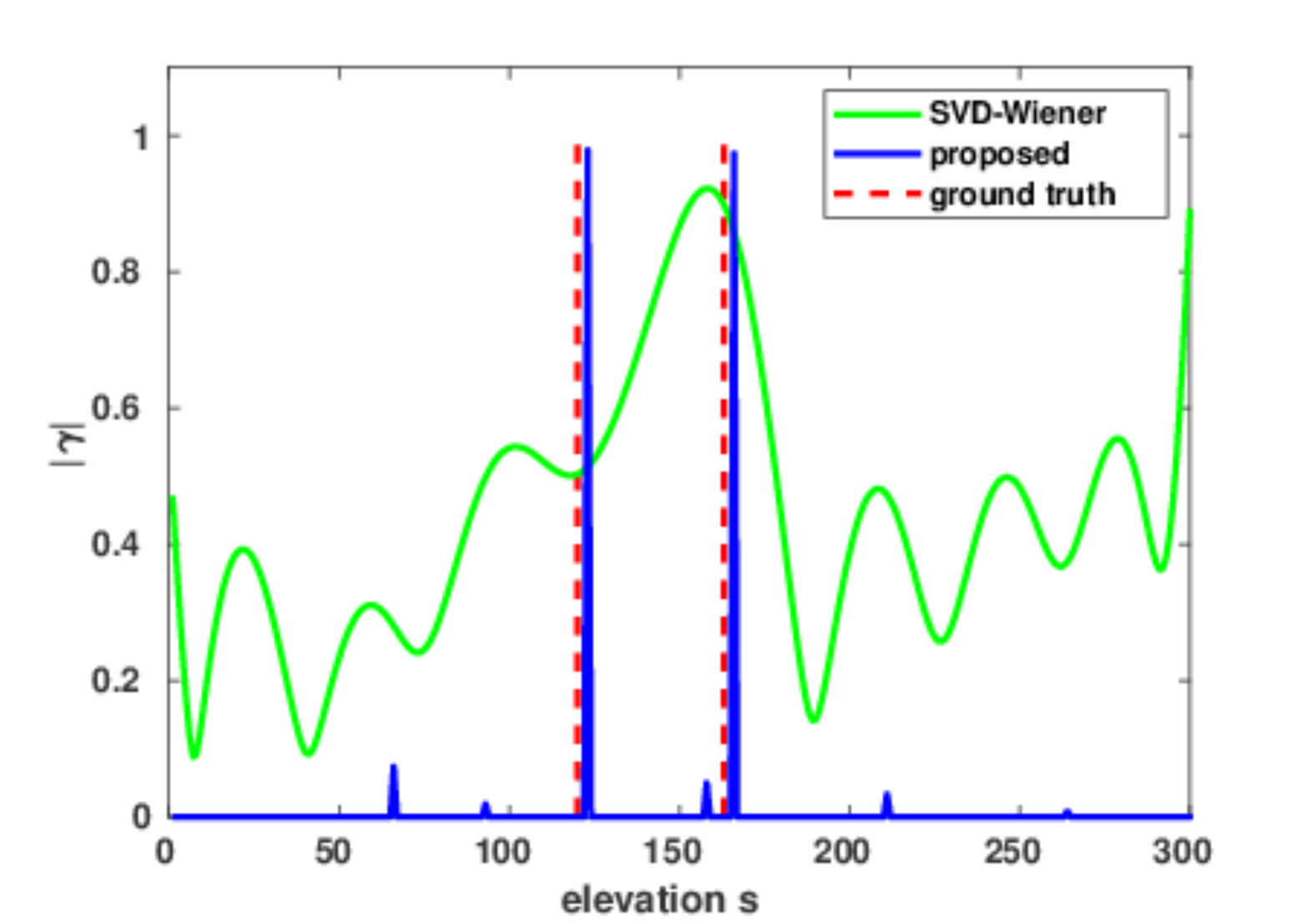}
    \caption*{(c)}
    \end{minipage}
    \begin{minipage}[t]{0.32\linewidth}
    \includegraphics[width=0.98\textwidth]{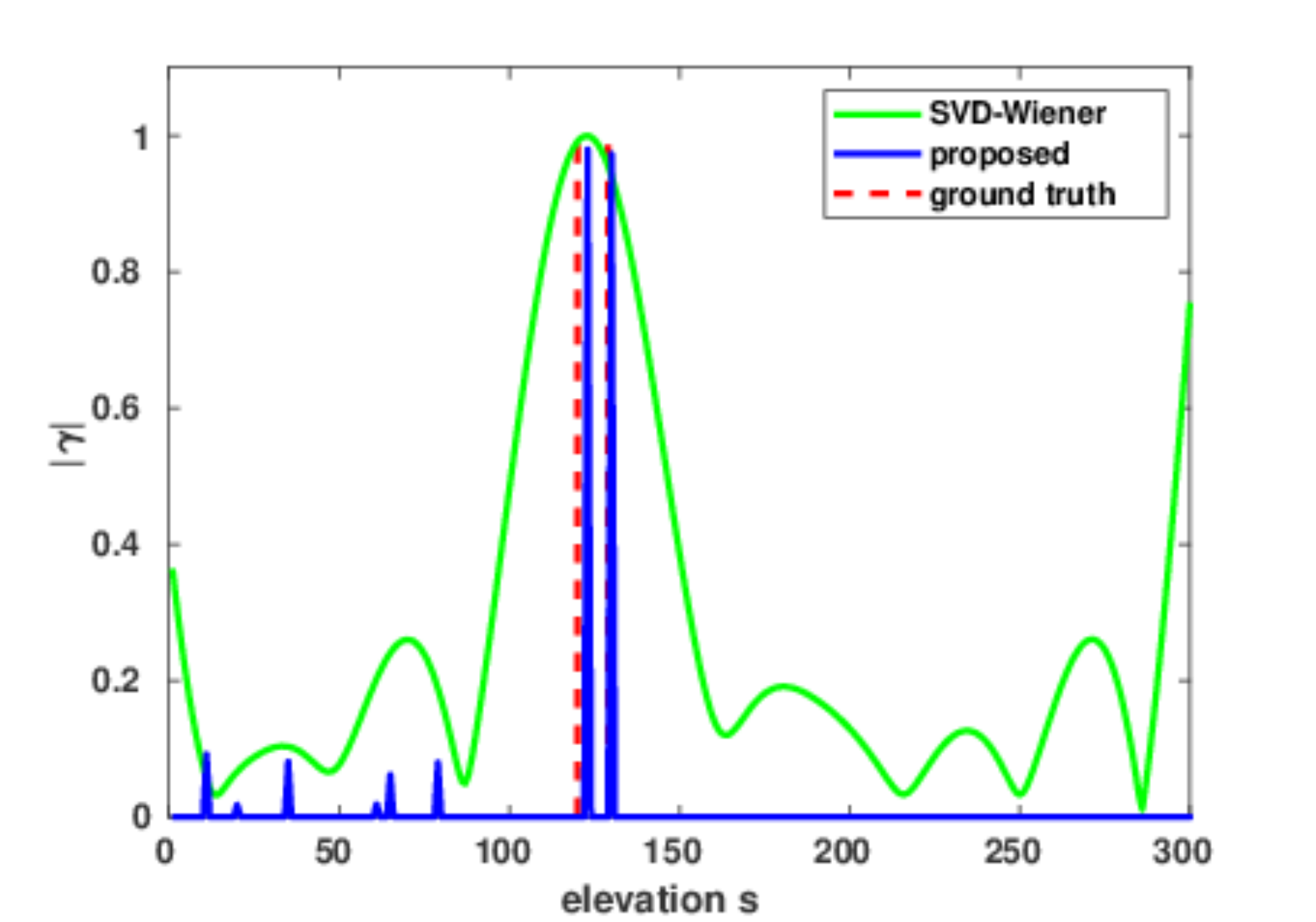}
    \caption*{(d)}
    \end{minipage}
    \begin{minipage}[t]{0.32\linewidth}
    \includegraphics[width=0.98\textwidth]{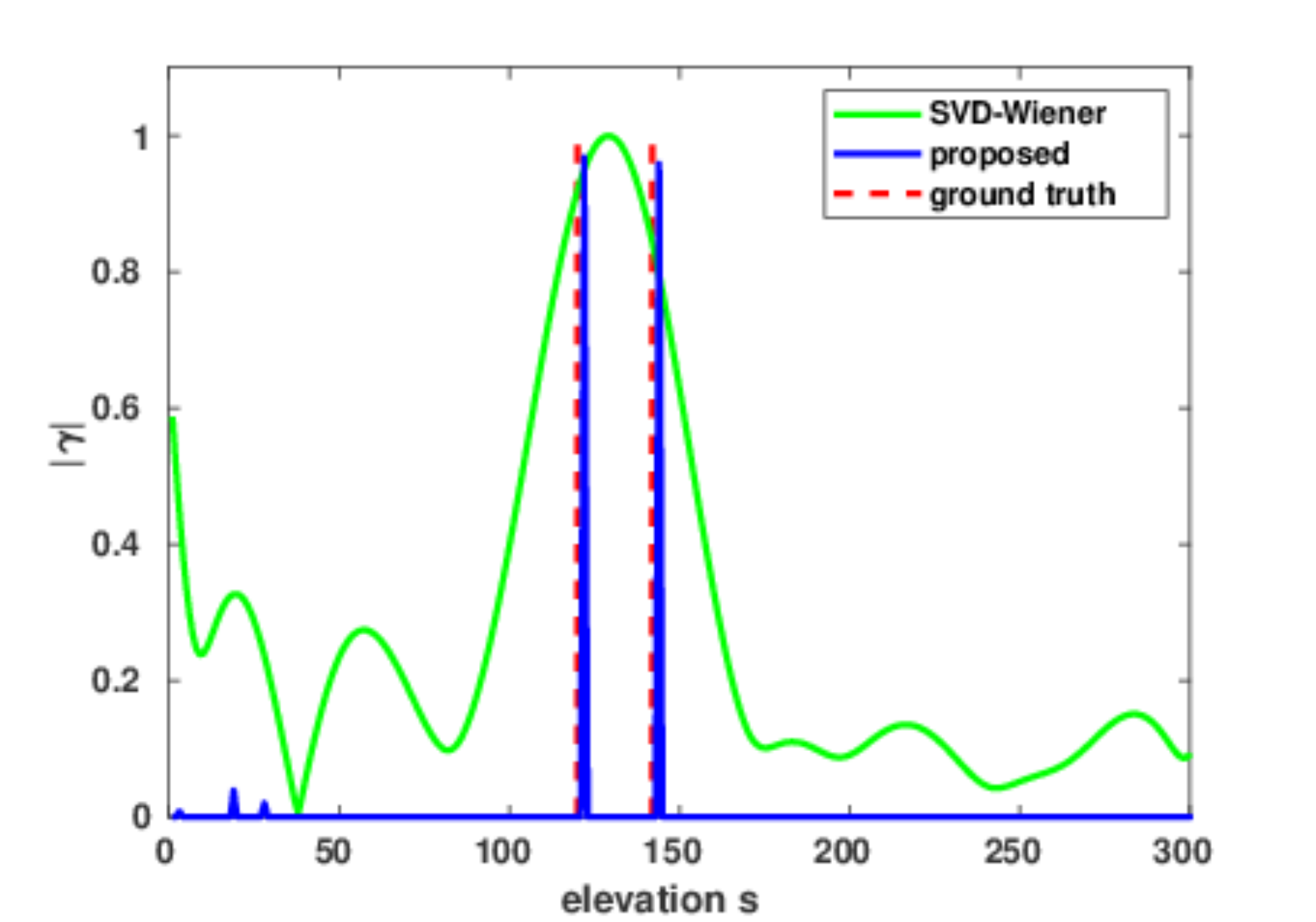}
    \caption*{(e)}
    \end{minipage}
    \begin{minipage}[t]{0.32\linewidth}
    \includegraphics[width=0.98\textwidth]{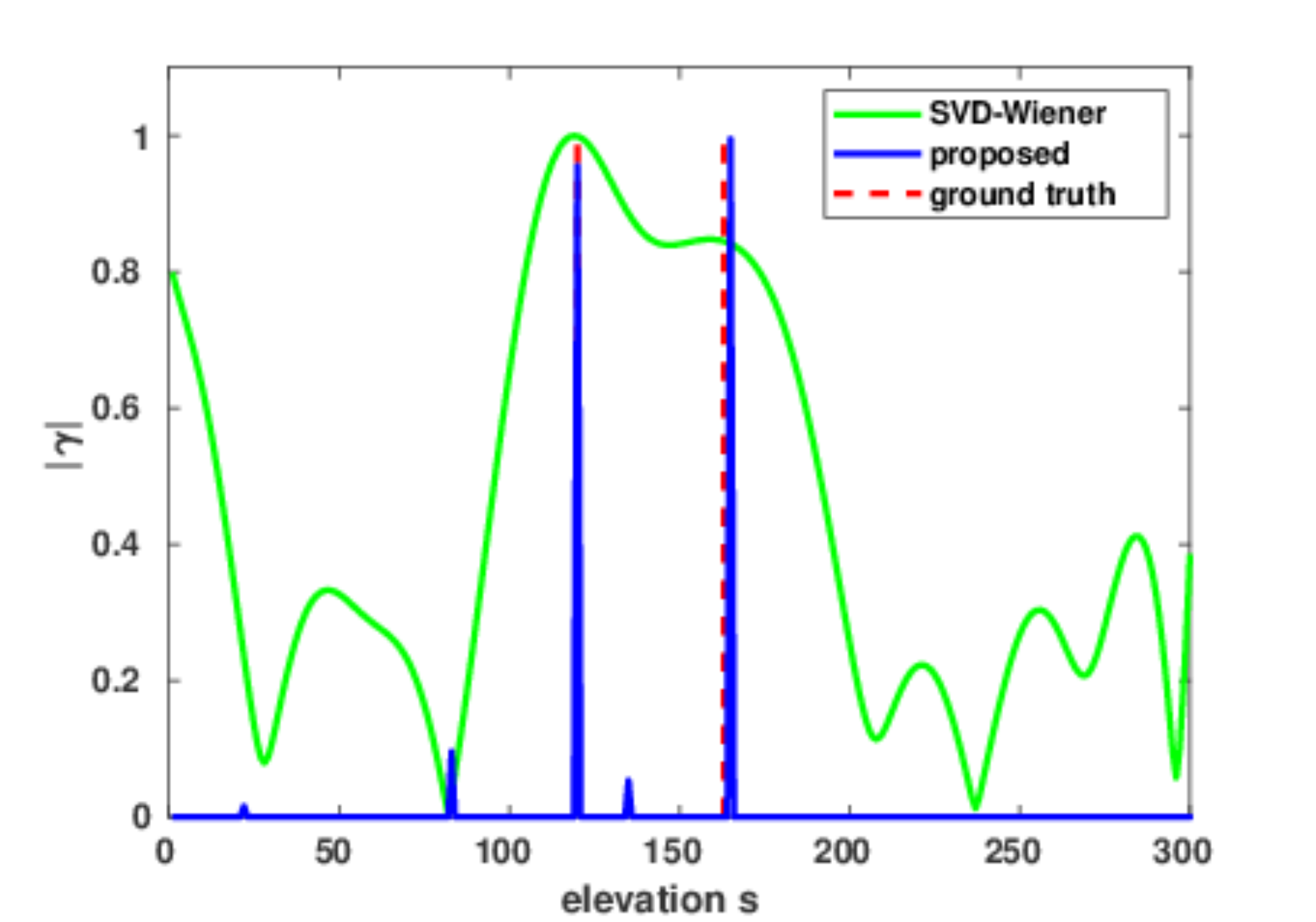}
    \caption*{(f)}
    \end{minipage}
    \caption{Estimated reflectivity profile of simulated data with overlaid double scatterers. From top to bottom, $SNR = 0,6$ dB. From left to right, the normalized elevation distance $\alpha= 0.2, 0.5, 1$.}
    \label{fig:comp_doub}
\end{figure*}
\begin{figure*}[h!]
	\centering
    \begin{minipage}[t]{0.49\linewidth}
    \includegraphics[width=0.98\textwidth]{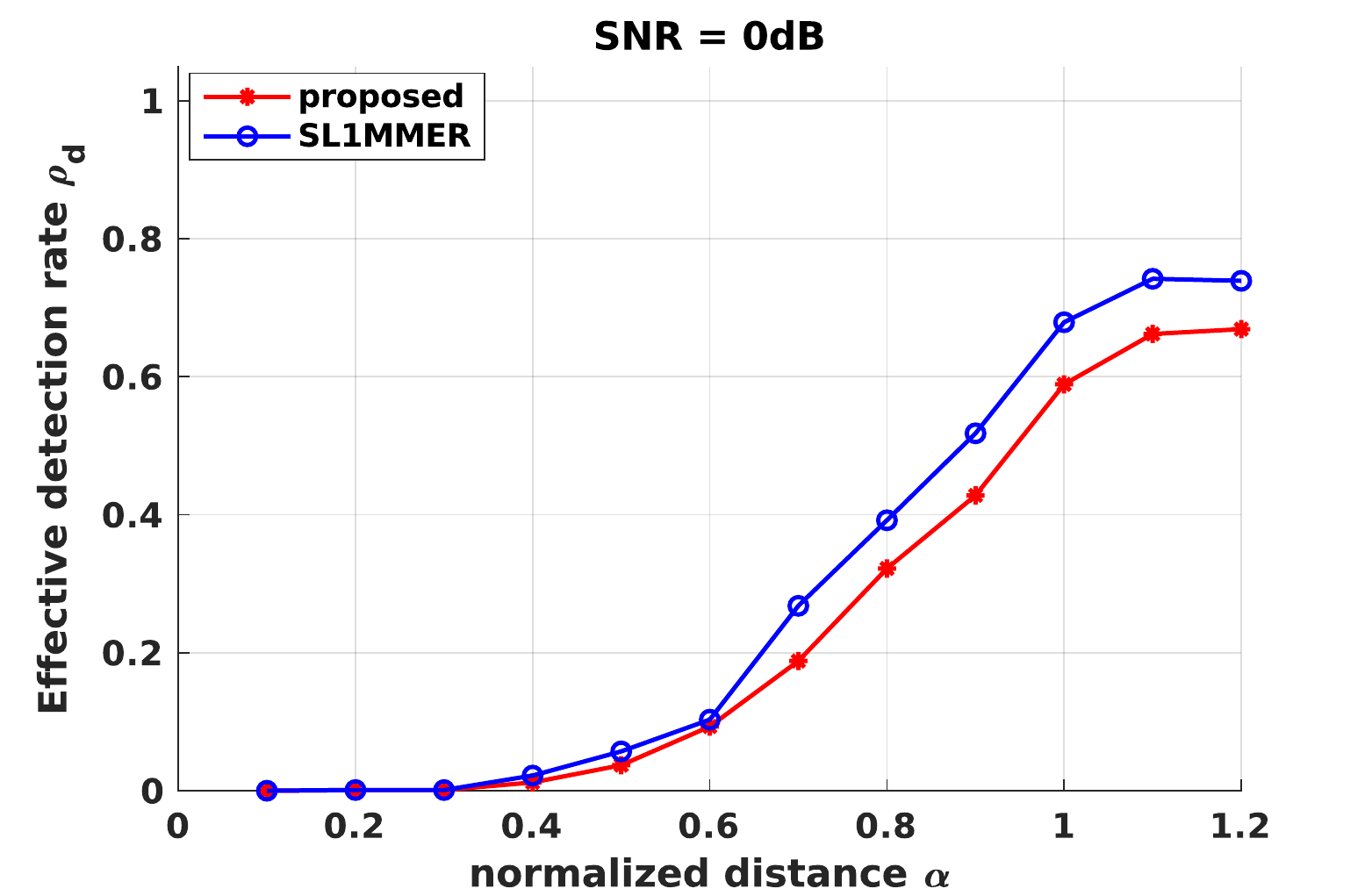}
    \caption*{(a)}
    \end{minipage}
    \begin{minipage}[t]{0.49\linewidth}
    \includegraphics[width=0.98\textwidth]{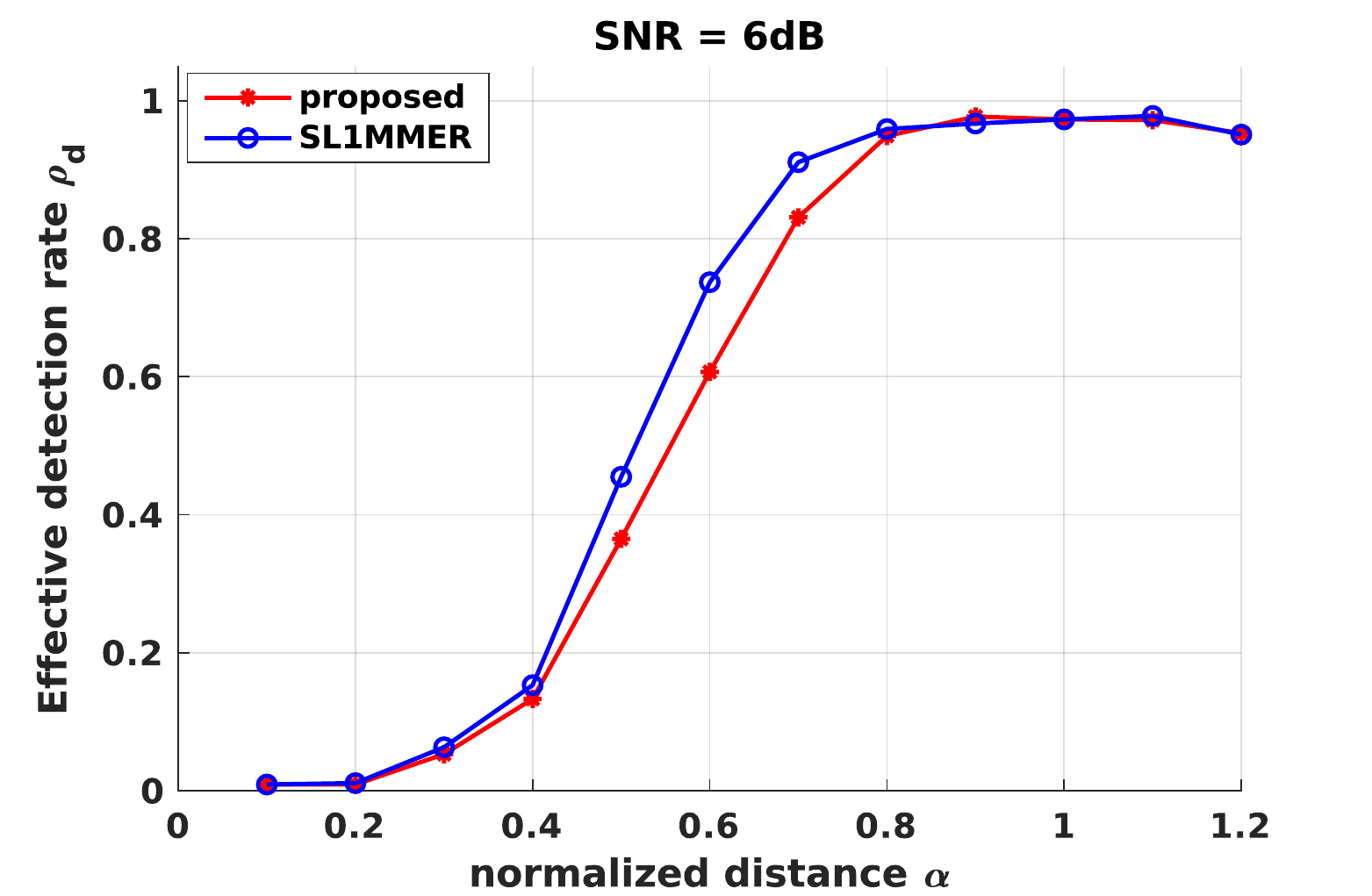}
    \caption*{(b)}
    \end{minipage}
    \caption{Detection rate $P_d$ as a function of the normalized elevation distance between the simulated facade and ground using the proposed algorithm (dashed star) and SL1MMER (dashed circle) with SNR = 0dB and 6dB, N = 25 and phase difference $\triangle \phi = 0$(worst case) under 0.2 million Monte Carlo trials.}
    \label{fig:detec_rate}
\end{figure*}

\begin{figure*}[h!]
    \centering
    \includegraphics[trim=2cm 0 1cm 0, clip, width=0.99\linewidth]{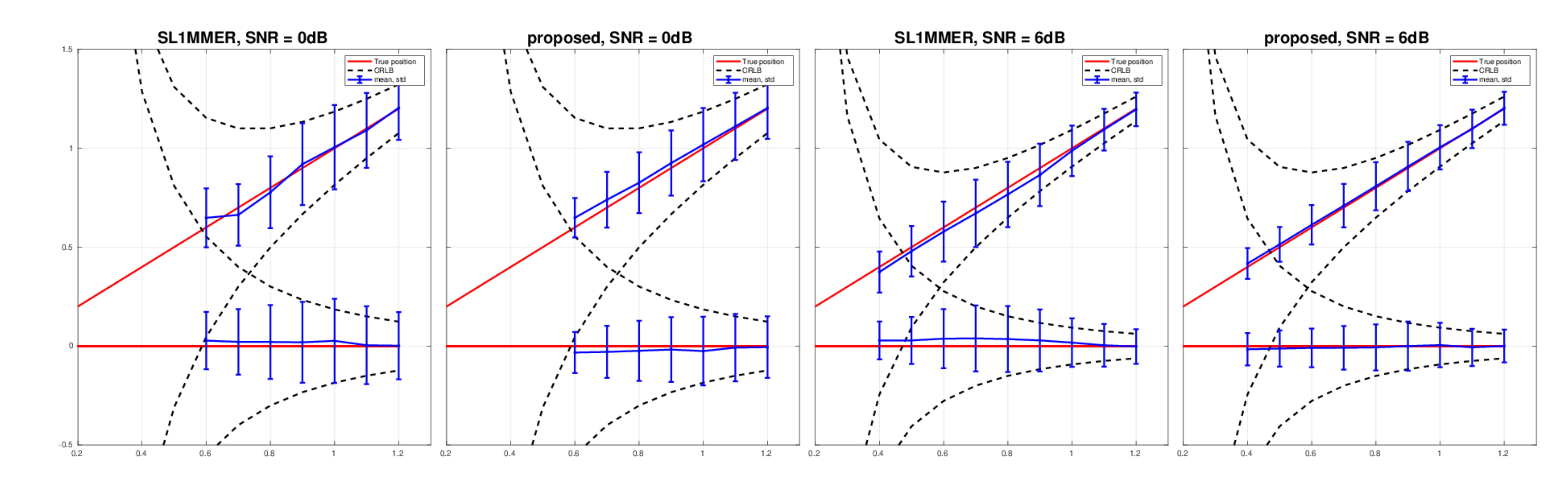}
    \caption{Estimated elevation of simulated facade and ground, (a) $SNR=0$dB with SL1MMER, (b) $SNR=0$dB with the proposed algorithm, (c) $SNR=6$dB with SL1MMER, (d) $SNR=6$dB with the proposed algorithm. Each dot has the sample mean of all estimates as its y value and the correspond standard deviation as error bar. The red line segments represent the true elevation of the simulated facade and ground. The dashed curves denote the true elevation $\pm 1 \times$CRLB normalized w.r.t the Rayleigh resolution.}
    \label{fig:fa_gr}
\end{figure*}

\begin{figure}[h]
    \centering
    \includegraphics[width=0.49\textwidth]{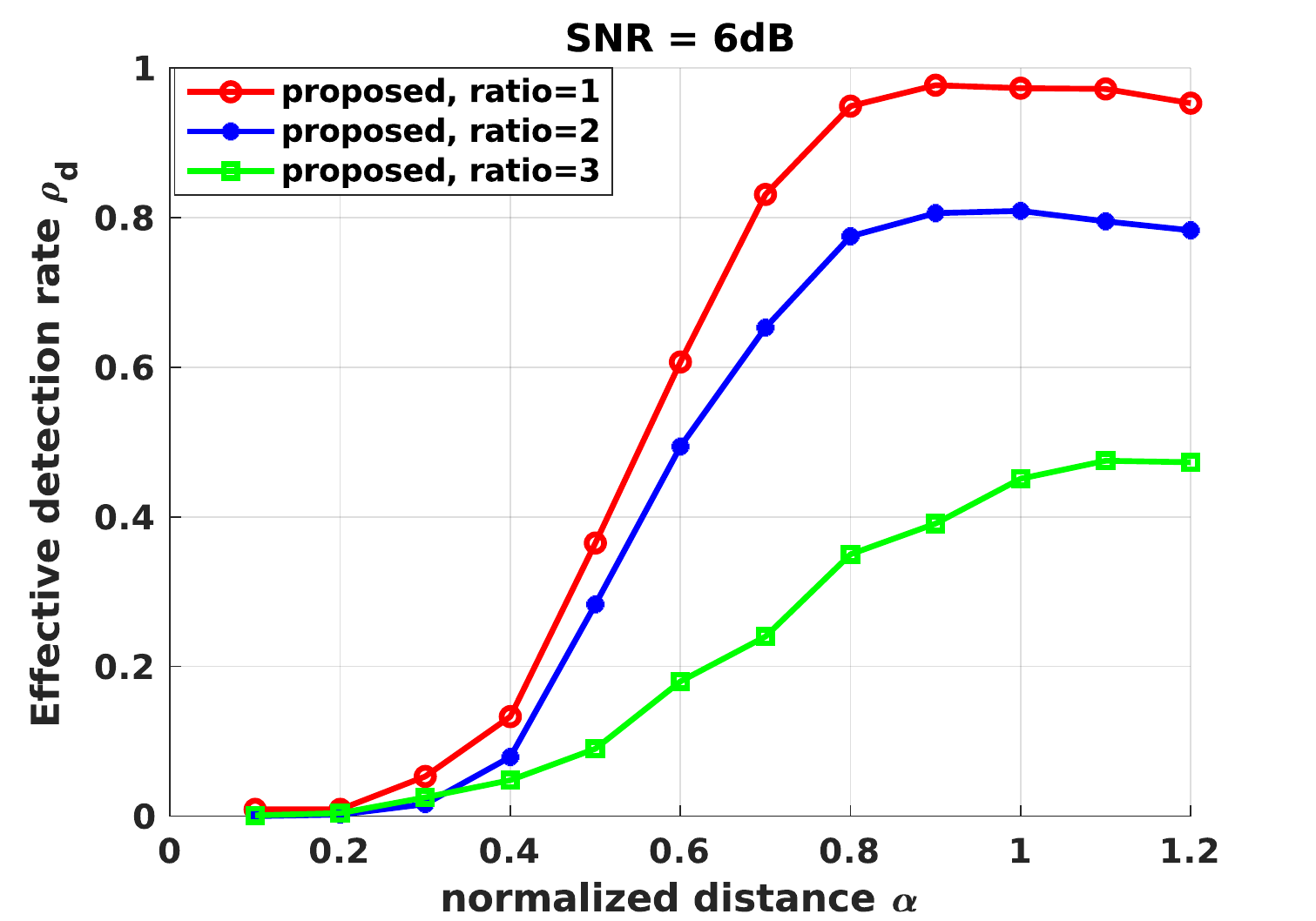}
    \caption{Effective detection rate $\rho_d$ as a function of amplitude ratio at $6$dB SNR.}
    \label{fig:dif_amp}
\end{figure}

\begin{figure}[h]
    \centering
    \includegraphics[width=0.49\textwidth]{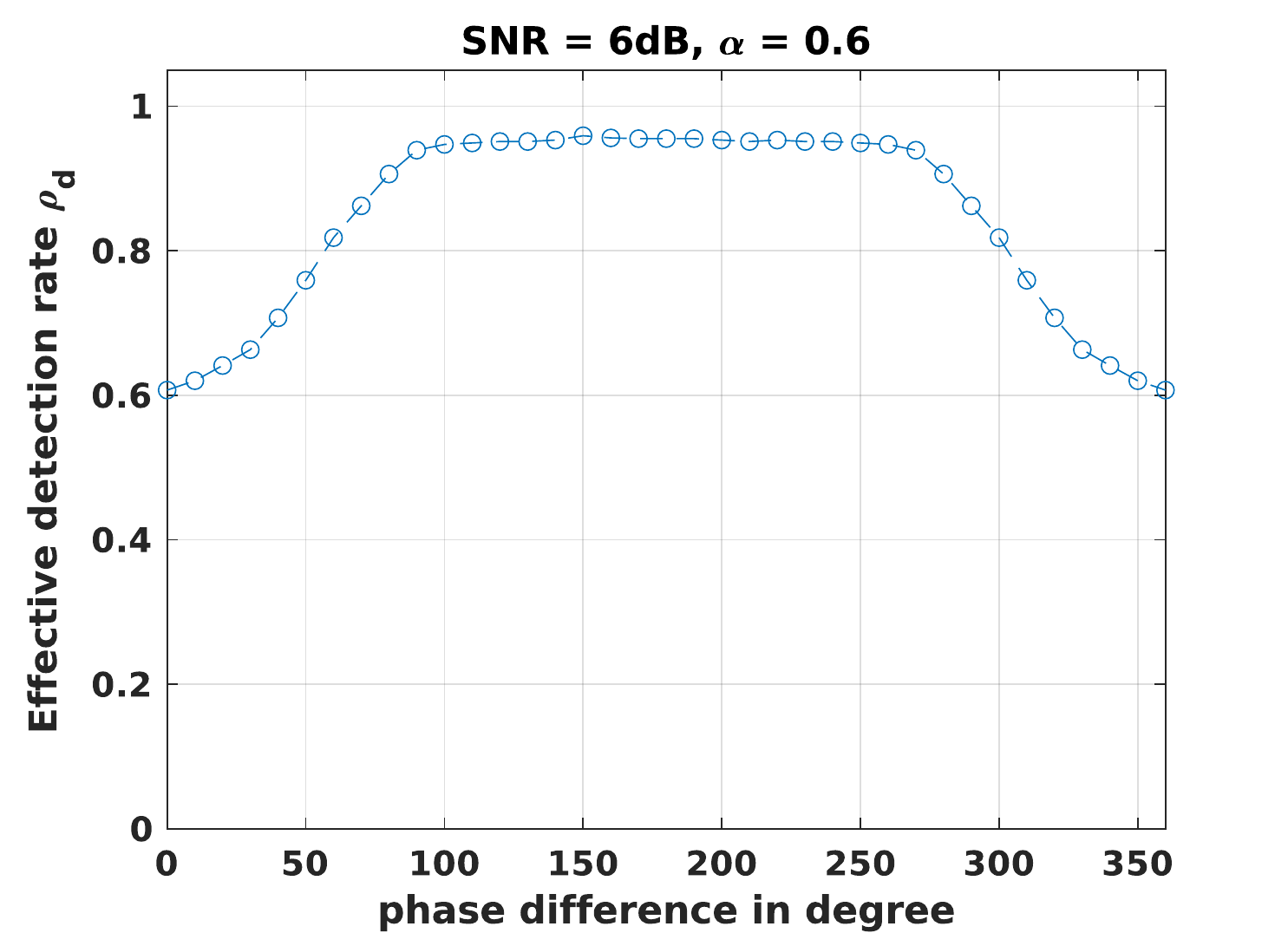}
    \caption{Effective detection rate $\rho_d$ as a function of phase difference $\triangle \phi$ under the case: $N=25$, $SNR=6$dB and $\alpha=0.6$.}
    \label{fig:vary_phase}
\end{figure}

For the double scatterer analysis, we simulate two-scatterer mixtures in the experiments and a systematic evaluation was carried out regarding the distance between simulated double scatterers, different scatterers amplitude ratio as well as phase difference between the double scatterers. 

\subsection*{\textbf{Performance with respect to scatterers distance}}
In this experiment, we performed a well-known TomoSAR benchmark test \cite{Zhu2010Tomographic} \cite{Zhu2010Very}. We simulated double scatterers with increasing elevation distance between the two layovered scatterers, in order to mimic a facade-ground interaction. Since we focus on the super-resolution regime, the elevation distance $d_s$ between the two overlaid scatterers is set to be no larger than 1.2 times of the Rayleigh resolution. Two different scenarios were taken into consideration with SNR$\in\{ 0, 6\}$ dB, which represent typical SNR levels in a high-resolution spaceborne SAR image. Fig. \ref{fig:comp_doub} demonstrates some examples of the estimated reflectivity profile at the normalized elevation distance $\alpha=[0.2,0.5,1.0]$. The normalized distance $\alpha$ is defined as the ratio of the elevation distance between the double scatterers and the Rayleigh resolution $\rho_s$, formally expressed as:
\begin{equation}
    \alpha = \frac{d_s}{\rho_s}
\end{equation}
From Fig. \ref{fig:comp_doub}, one can see that both the trained $\boldsymbol{\gamma}$-Net and SVD-Wiener are able to distinguish the overlaid double scatterers in the non-superresolving case, i.e. last column, when $\alpha=1.0$. But comparing to SVD-Wiener, $\boldsymbol{\gamma}$-Net provides much higher elevation estimation accuracy. Moreover, when we move the double scatterers closer into the Rayleigh resolution, SVD-Wiener fails to separate them. In the contrast, $\boldsymbol{\gamma}$-Net is still capable of detecting the double scatterers in most cases, which exhibits its super-resolution power. 

Hereafter, we compare the proposed algorithm with the state-of-the-art SL1MMER algorithm \cite{Zhu2012Super-Resolution} focusing on the detection rate and the estimation accuracy.
Similar to the single scatterer case, we use the \textbf{\textit{effective detection rate}} to fairly evaluate the detection rate. An effective detection of double scatterer is defined as: 
\begin{enumerate}
    \item the hypothesis test correctly decides two scatterers for a double-scatterers signal;
    \item the estimated elevation of \textbf{\textit{both}} detected double scatterers are within $\pm3$ times CRLB w.r.t their true elevation;
    \item both elevation estimates are also within $\pm0.5 \ d_s$ w.r.t their true elevation.
\end{enumerate}
The third criterion is seldom seen in the literature. However, it is necessary, because in extremely super-resolving cases, 3 times CRLB will become much larger than the elevation distance. Hence, it cannot be used as a accountable measure for reasonable estimates. $\pm0.5 \ d_s$ is a much stricter constraint in such cases, which will reflect the true performance of the algorithm.

Fig. \ref{fig:detec_rate} compares the effective detection rate $P_d$ of SL1MMER and the proposed algorithm for the case $N=25$. For each pair of ($SNR,\alpha$), 0.2 million Monte Carlo trials for the worst case in TomoSAR inversion, i.e. the double scatterers have the same amplitude and phase, were simulated. The effective detection rate $P_d$ is presented as a function of the normalized distance. The red and blue polylines illustrate the results of the proposed algorithm and SL1MMER, respectively. As we can see from Fig. \ref{fig:detec_rate}, the proposed algorithm has comparable super-resolution power as SL1MMER. 

Fig. \ref{fig:fa_gr} demonstrates the elevation estimates of simulated facade and ground w.r.t the true normalized elevation distance. In each subplot of Figure \ref{fig:fa_gr}, the two red line segments represent the true elevation of the simulated facade and ground, respectively, while the dashed lines show the true elevation $\pm 1 \times$CRLB (normalized). Exhaustive details of the derivation of the CRLB can be found in \cite{Zhu2012Super-Resolution}. The elevation estimates of simulated facade and ground are plotted with each dot depicting the sample mean of all estimates at the given normalized distance and the error bar indicating the corresponding standard deviation. Points below an effective detection rate of 10\% were not plotted in the figure. As it is shown in Fig. \ref{fig:fa_gr}, the proposed algorithm shows higher elevation estimation accuracy than SL1MMER. To be specific, at 0dB SNR, although both the proposed algorithm and SL1MMER have similar estimate bias, the proposed algorithm leads to much smaller variance. In high SNR case, the proposed algorithm outperforms SL1MMER in super-resolving cases w.r.t the elevation estimation accuracy. As can be seen that, SL1MMER suffers from much larger elevation estimate bias as well as the standard deviation. 

\subsection*{\textbf{Performance with respect to amplitude ratio}}
This simulation sets out to evaluate the performance of the proposed algorithm w.r.t different amplitude ratio of the double scatterers. Fig \ref{fig:dif_amp} illustrates us the effective detection rate of the proposed algorithm at different amplitude ratio. As can be seen, the effective detection rate decreases with the increasing amplitude ratio. Since $\boldsymbol{\gamma}$-Net promotes sparsity by shrinking elements with small magnitude layer by layer. With the increase of the amplitude ratio between simulated double scatterers, the darker scatterer becomes less prominent, and hence easier to be ignored. Therefore, at high amplitude ratio, the proposed algorithm tends to detect single scatterer with dominant amplitude. However, from our perspective, it will not affect the application of the proposed algorithm. In real-world processing, if one scatterer is much more prominent than others in a pixel, we can usually judge that this pixel contains only a single scatterer by viewing others as noise.

\subsection*{\textbf{Performance with respect to phase difference}}
We varied the phase difference between the simulated double scatterers in this simulation to further verify the generalization ability of the proposed algorithm. Fig. \ref{fig:vary_phase} demonstrates us an example of the effective detection rate when $N=25$ and $SNR=6$dB, with the normalized distance $\alpha=0.6$. The double scatterers in the simulation are set to have identical amplitude.  As we can see, although the phase difference $\triangle \phi$ affects the performance, the proposed algorithm is still capable of providing satisfactory super-resolution power even in the worst case, where the phase difference $\triangle \phi = 0$. 


\subsection{Analysis of false detection}
In this section, we will provide a quantitative assessment about false detection. We used the proposed algorithm to detect 0.2 million samples containing 0 scatterer, i.e. pure noise, at different SNR. As it is shown in Table \ref{table:noise_eva}, the proposed algorithm is able to distinguish almost all samples of noise at different SNR. Less than 5 percent samples are falsely detected as single scatterer and only about $0.1 \%$ as double scatterers. The low false alarm attributes to the powerful model order selection with known noise variance in the simulation. However, in real-world application, the noise variance needs to be estimated. Therefore, Table \ref{table:noise_eva} shows the upper limit.
\begin{table}[h]
\centering
\begin{tabular}{p{0.2\textwidth}| p{0.2\textwidth}}
\toprule
pure noise detected as & percentage \\
\midrule
0 scatterer & $95.57 \%$ \\
1 scatterer & $4.33 \%$ \\
2 scatterers & $0.1 \%$ \\
\bottomrule
\end{tabular}
\caption{Statistics of evaluation on false detection.} 
\label{table:noise_eva}
\end{table}

\subsection{Performance at limited number of measurements}
\begin{figure*}[h!]
	\centering
    \begin{minipage}[t]{0.49\linewidth}
    \includegraphics[height=6cm]{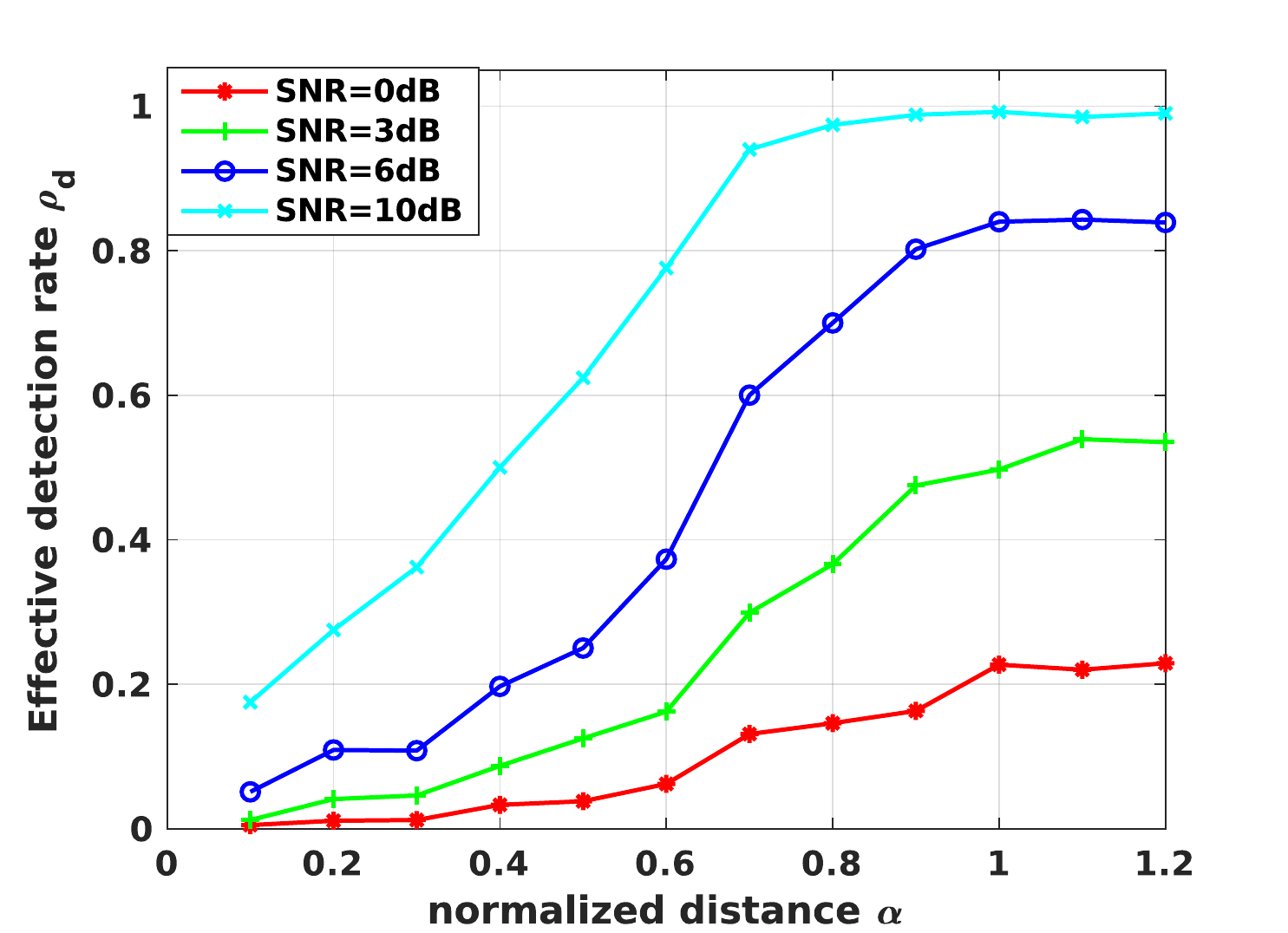}
    \caption*{(a) proposed}
    \end{minipage}
    \begin{minipage}[t]{0.49\linewidth}
    \includegraphics[height=6cm]{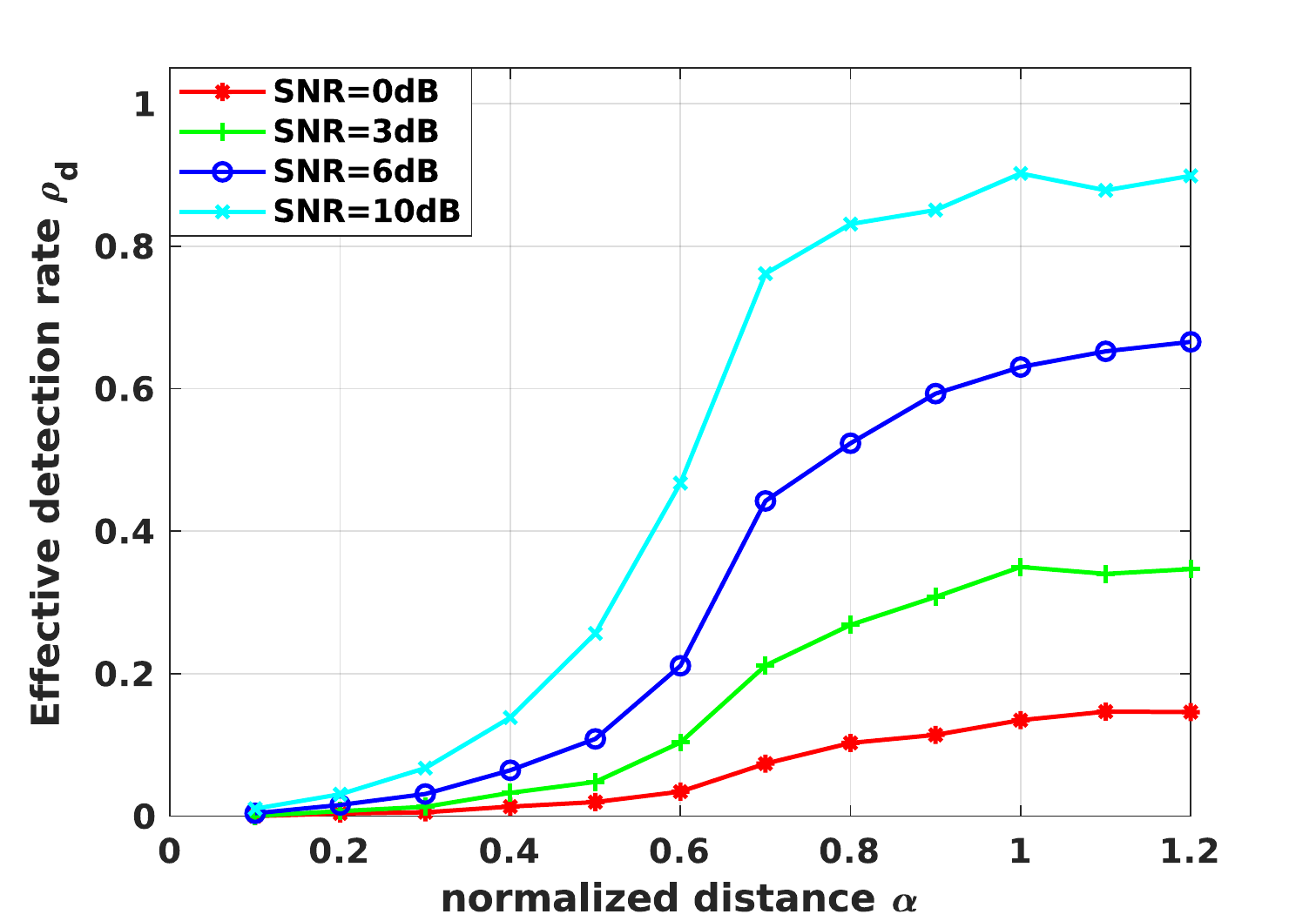}
    \caption*{(b) SL1MMER}
    \end{minipage}
    \caption{Effective detection rate $P_d$ as a function of the normalized elevation distance between double scatterers simulated with 6 real baselines. The simulated double scatterers are set to have identical phase and amplitude, i.e. the worst case. For each pair of (SNR, $\alpha$), 0.2 million Monte Carlo trials were simulated. (a) the proposed algorithm, (b) SL1MMER}
    \label{fig:comp_PM7}
\end{figure*}

This simulation was carried out to verify the performance of the proposed algorithm at limited number of baselines. We simulated data with only 6 baselines according to a real TanDEM-X images stack we have. The baseline ranges from -565.5m to 373.2m. Fig. \ref{fig:comp_PM7} compares the performance of the two algorithms at limited number of measurements. As one can see, in the noisy case, i.e. SNR = 0dB, the two algorithms have similar performance. However, with the increase of the SNR level, the proposed algorithm outperforms SL1MMER by a fair margin. 

\subsection{Performance verification using real data}
\begin{figure*}[h]
    \centering
    \includegraphics[width=0.95\linewidth]{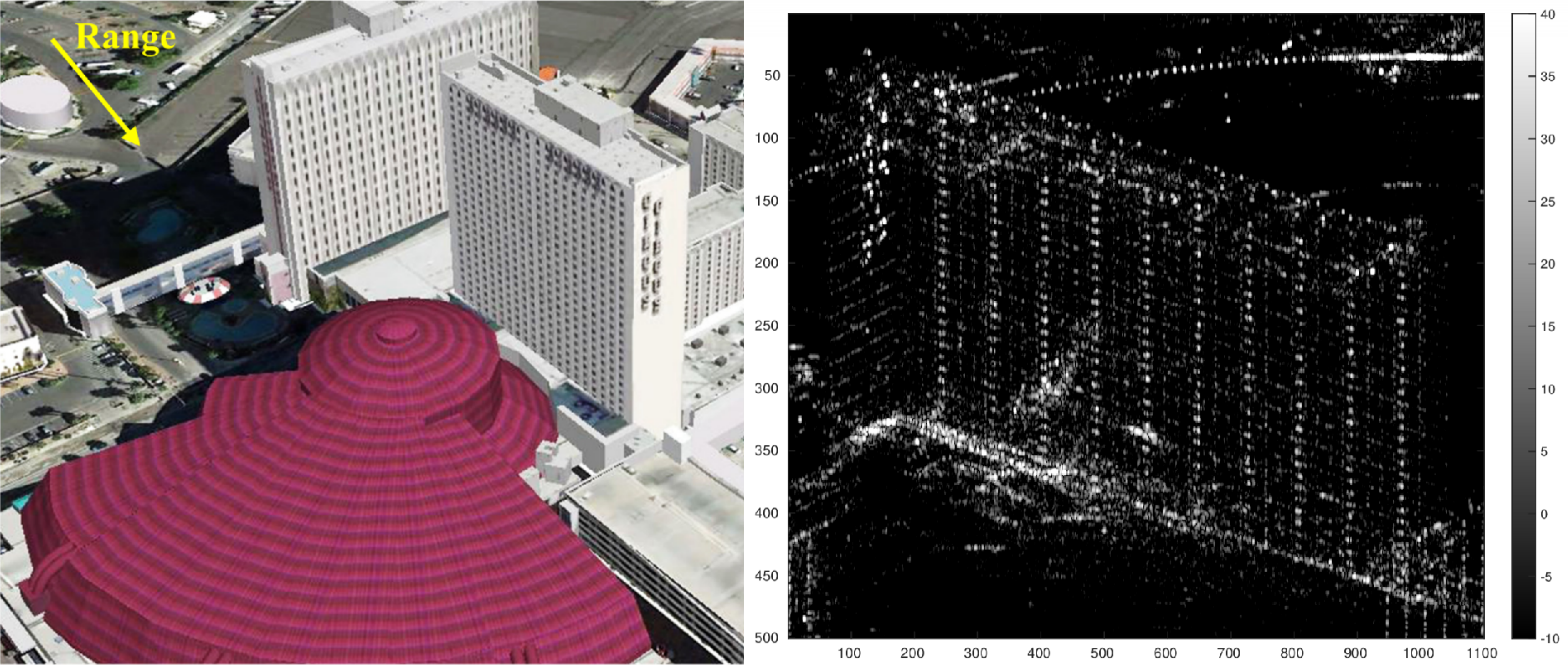}
    \caption{Test site. Left: optical image from Google Earth, right: SAR mean intensity image}
    \label{fig:test_are}
\end{figure*}

For a better evaluation of the proposed algorithm, we worked with a stack of six high-resolution TanDEM-X pairs acquired in the bistatic mode mentioned in the previous section. The elevation aperture size of about 940m results in about 12m inherent elevation resolution. An optical image of the test site from Google Earth and the SAR mean intensity image are showed in Fig \ref{fig:test_are}. The yellow arrow indicates the range direction. The atmospheric effects and deformation are ignored since the temporal baselines are negligible. Preprocessing, such as multiple SAR images co-registration and phase calibration were carried out using the DLR's integrated wide area processor (IWAP) \cite{rodriguez_gonzalez_integrated_2013}. Moreover, we manually selected a coherence point on the ground as reference and set its elevation as zero. 

$\boldsymbol{\gamma}$-Net employed for the real data experiment was trained with data simulated using the real baseline distribution. The training data contains 4 million samples generated with the same strategy mentioned in the simulation setup. After training, $\boldsymbol{\gamma}$-Net can be directly applied in the upcoming TomoSAR processing on real data.

\begin{figure*}[h]
    \centering
	\begin{minipage}[t]{0.49\linewidth}
	    \includegraphics[width=\textwidth]{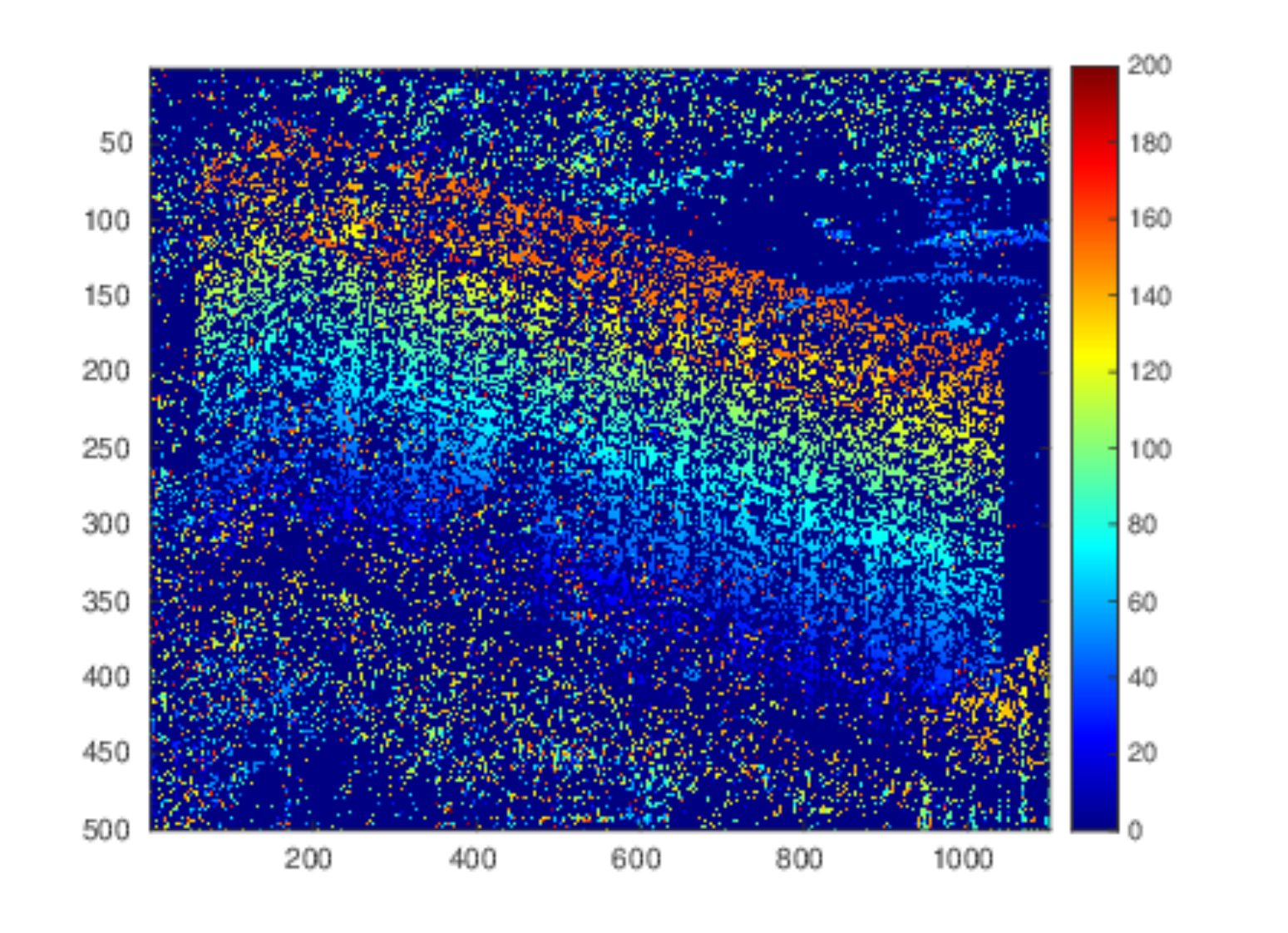}
	    \caption*{(a)}
	\end{minipage}
	\begin{minipage}[t]{0.49\linewidth}
		\includegraphics[width=\textwidth]{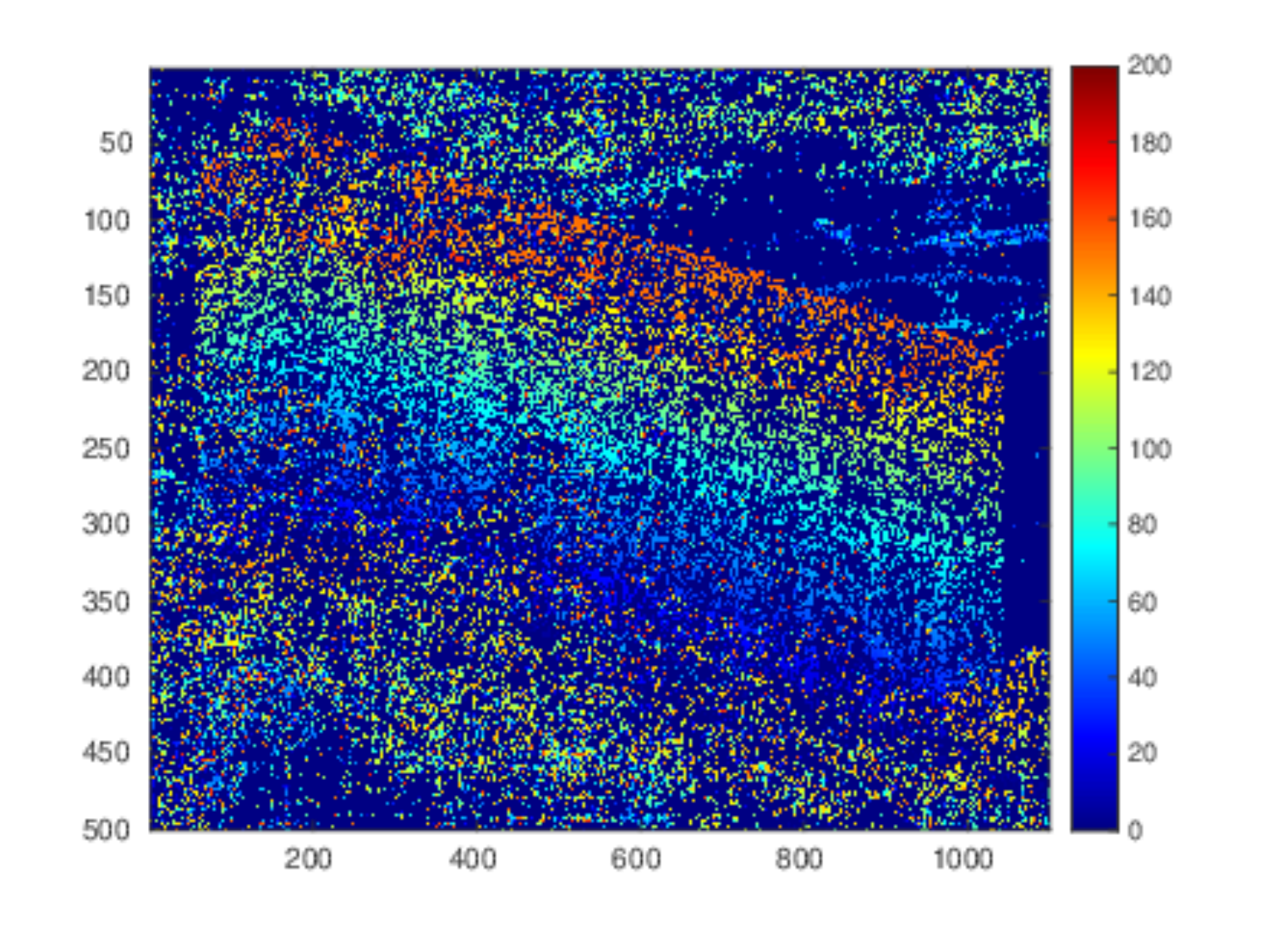}
		\caption*{(b)}
	\end{minipage}
    \caption{Reconstructed and color-coded elevation of detected single scatterer. (a) the proposed algorithm, (b) the SL1MMER algorithm}
    \label{fig:sin}
\end{figure*}
\begin{figure*}[h]
    \centering
	\begin{minipage}[t]{0.49\linewidth}
		\includegraphics[width=\textwidth]{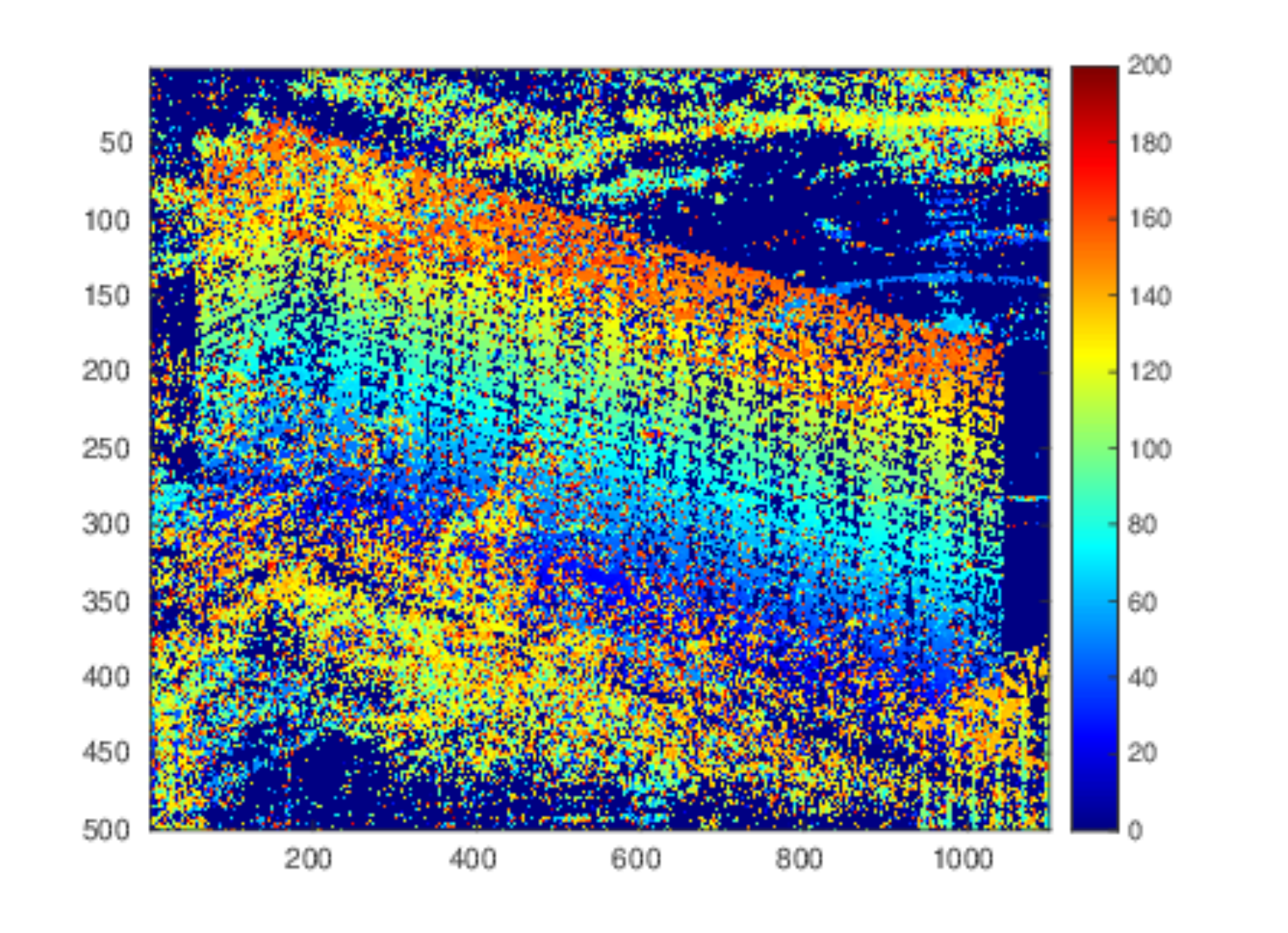}
		\caption*{(a)}
	\end{minipage}
	\begin{minipage}[t]{0.49\linewidth}
		\includegraphics[width=\textwidth]{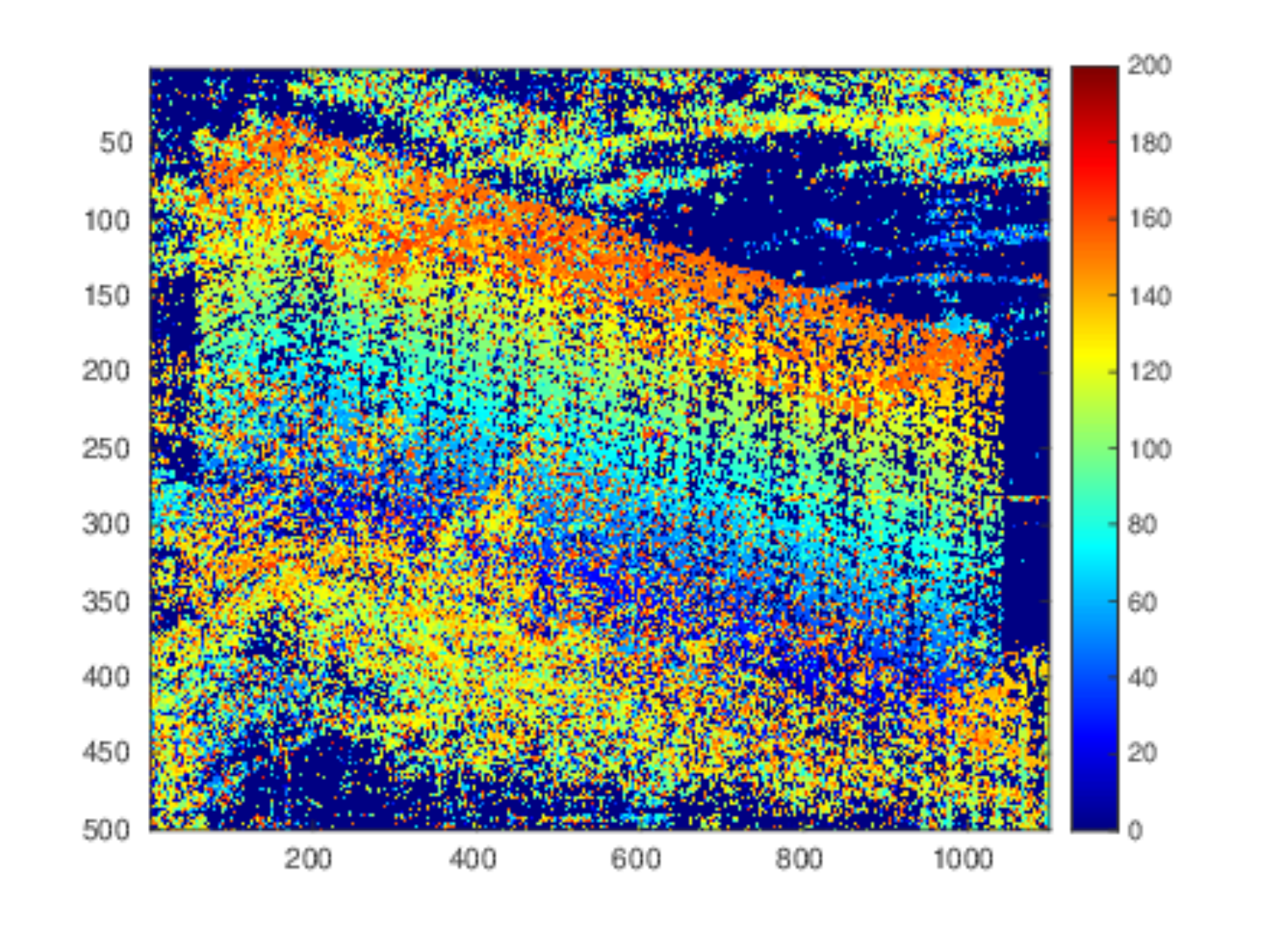}
		\caption*{(b)}
	\end{minipage}
	\caption{Reconstructed and color-coded elevation of detected single scatterer $+$ top layer of the detected double scatterers. (a) the proposed algorithm, (b) the SL1MMER algorithm.}
    \label{fig:sin+top}
\end{figure*}
\begin{figure*}[h]
    \centering
	\begin{minipage}[t]{0.49\linewidth}
		\includegraphics[width=\textwidth]{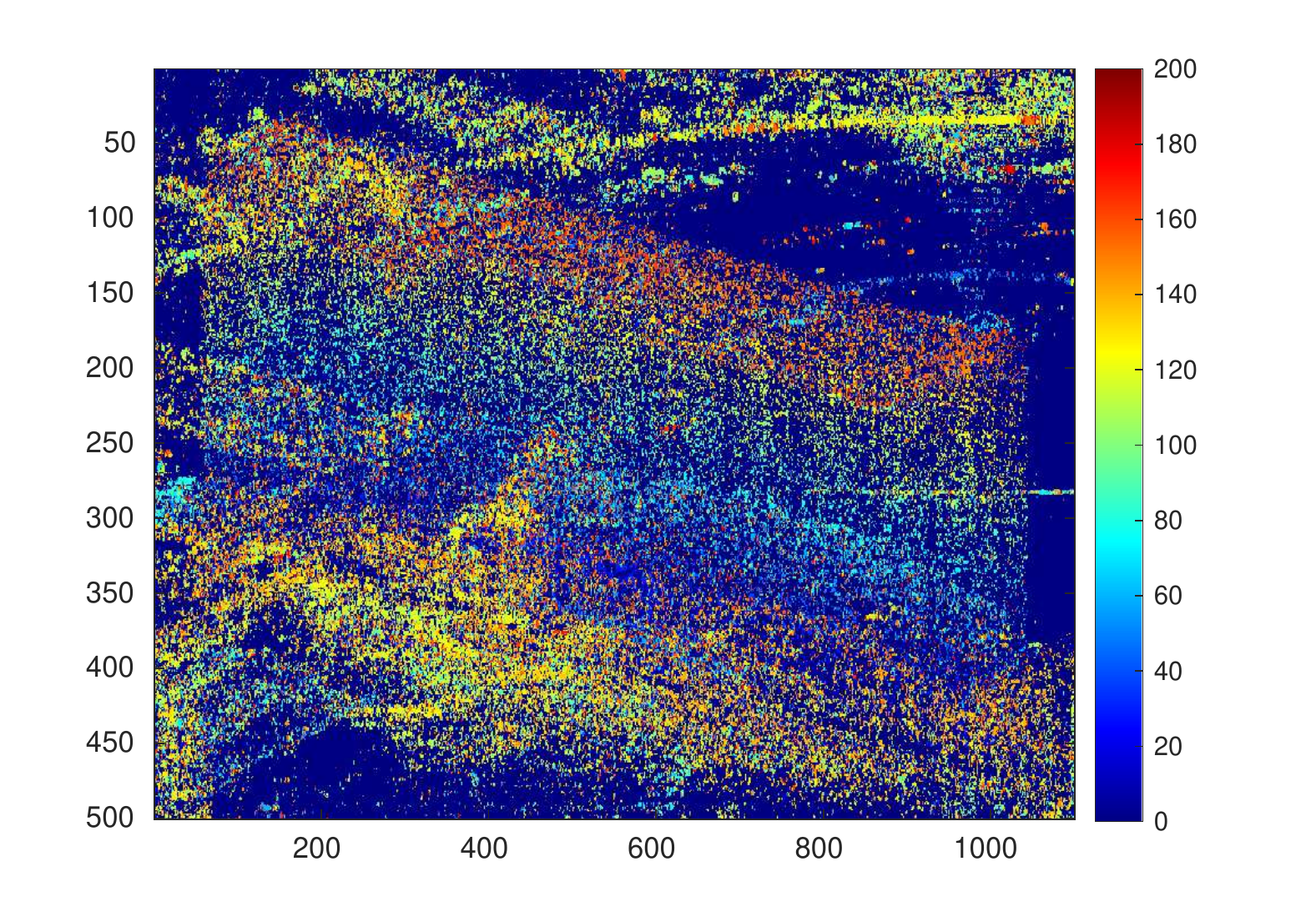}
		\caption*{(a)}
	\end{minipage}
	\begin{minipage}[t]{0.49\linewidth}
		\includegraphics[width=\textwidth]{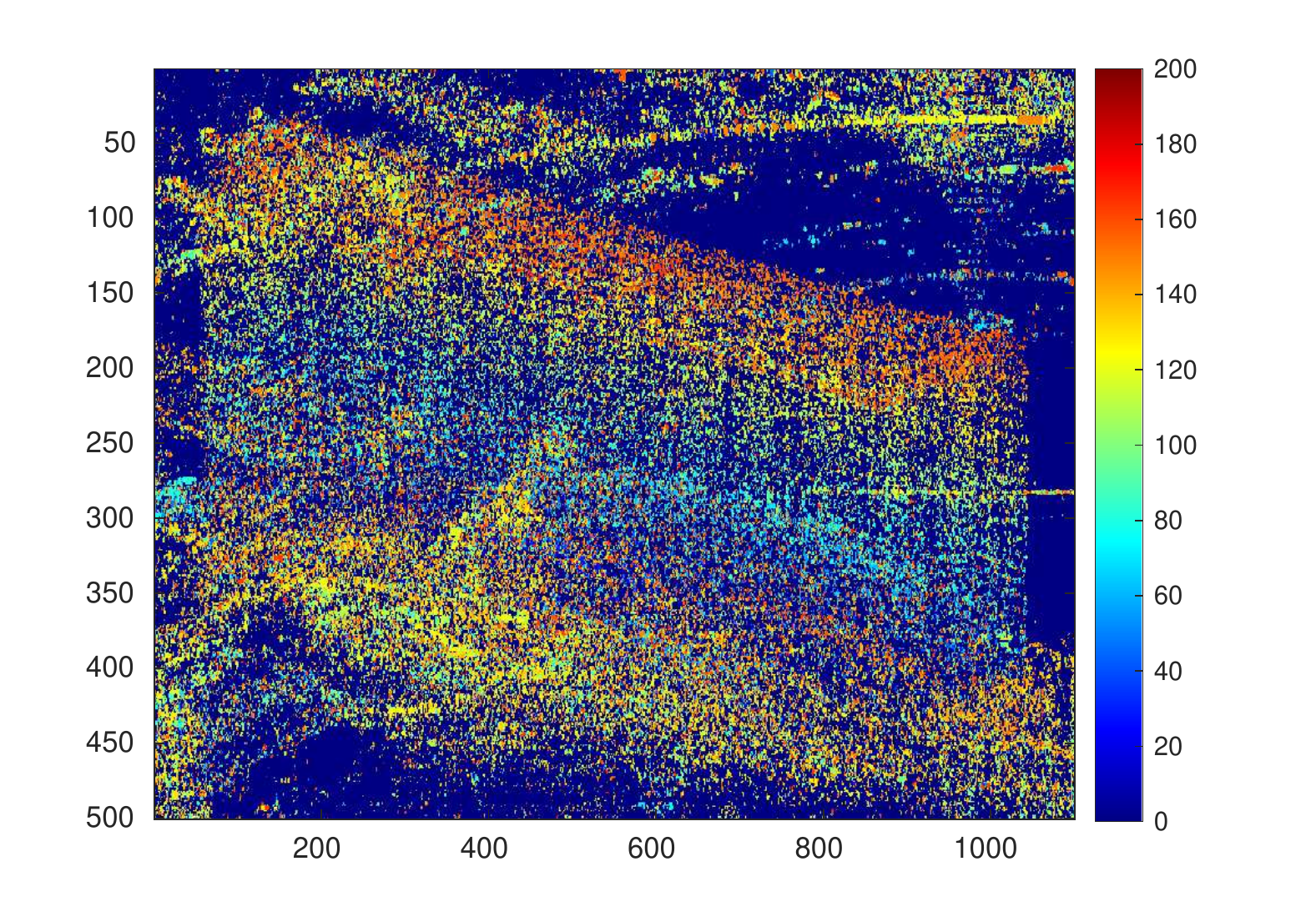}
		\caption*{(b)}
	\end{minipage}
	\begin{minipage}[t]{0.49\linewidth}
		\includegraphics[width=\textwidth]{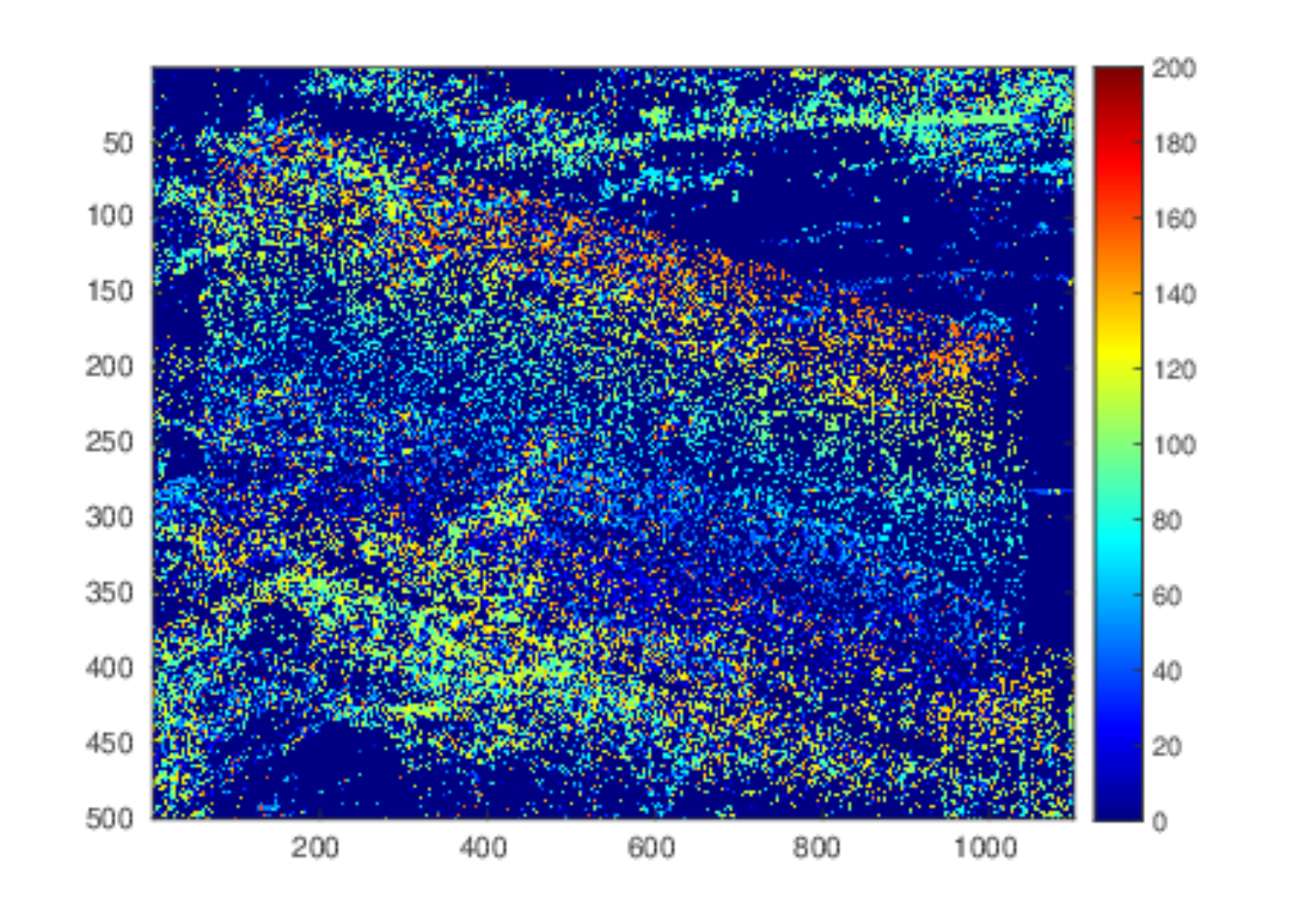}
		\caption*{(c)}
	\end{minipage}
	\begin{minipage}[t]{0.49\linewidth}
		\includegraphics[width=\textwidth]{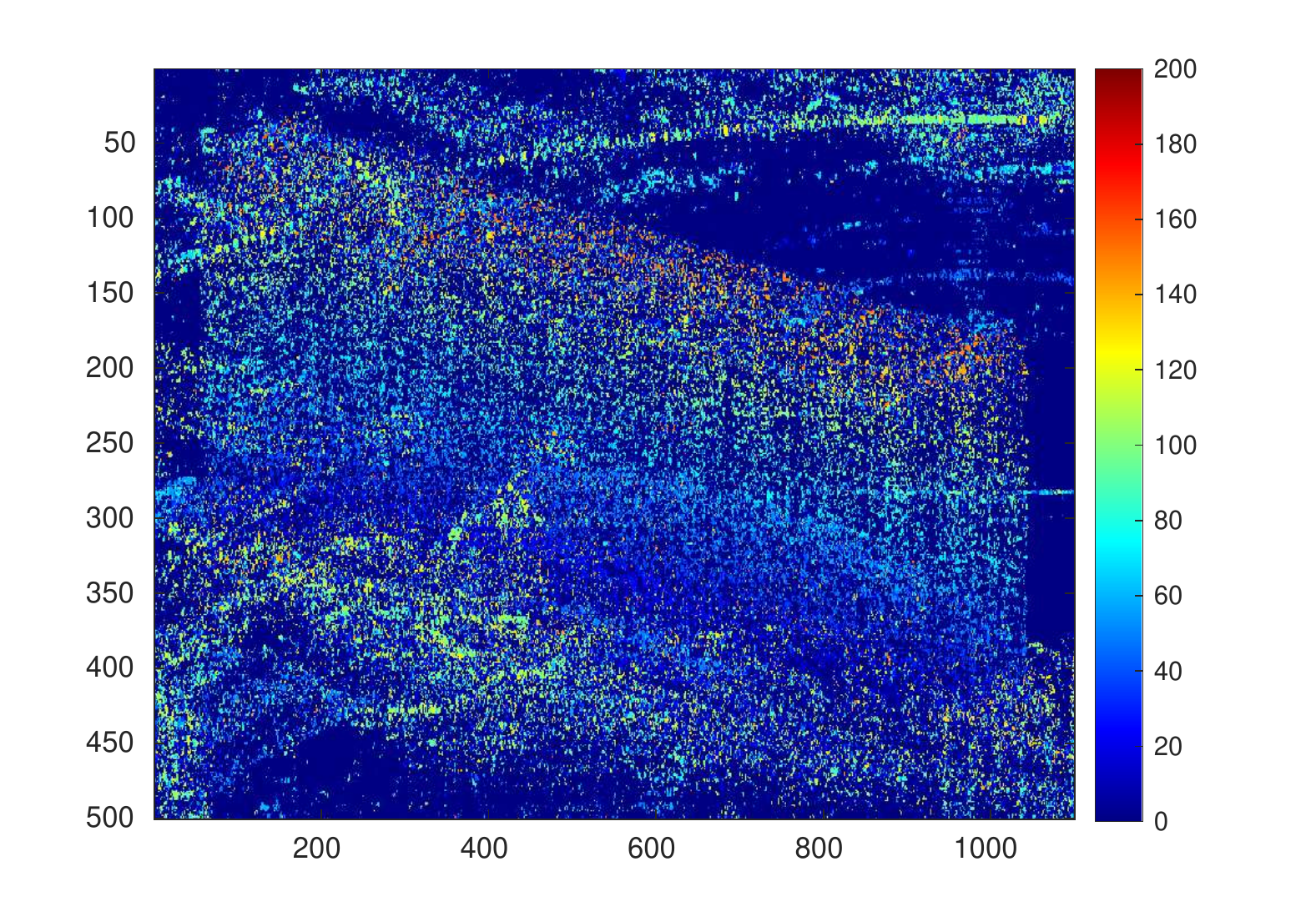}
		\caption*{(d)}
	\end{minipage}
	\caption{Reconstructed and color-coded elevation of detected double scatterers. From top to down: top and bottom layer, respectively. From left to right: the proposed and SL1MMER algorithm, respectively.}
    \label{fig:top+bot}
\end{figure*}

We use the proposed algorithm to reconstruct the whole test site and demonstrate the super-resolution power by comparing to the results derived by SL1MMER. The complete comparison of the reconstruction results of the test site are demonstrated in Figs \ref{fig:sin}-\ref{fig:top+bot}. Fig \ref{fig:sin} depicts only the reconstruction of single scatterer detected by both algorithms. Fig \ref{fig:sin+top} demonstrates the elevation estimates of detected single scatterers combined with the top layer of detected double scatterers. Comparing Fig. \ref{fig:sin+top} to Fig. \ref{fig:sin}, we can see that both algorithms are able to detect dense double scatterers,which contribute to significant information increment and complete the structure of individual buildings shown in the test site. For a better view of the separation of overlaid scatterers, we demonstrate the top and bottom layer of detected double scatterers separately in Fig \ref{fig:top+bot}. As we can see, the density of double scatterers detected by the proposed algorithm and SL1MMER is almost the same, meaning that both algorithms possess similar or even the same super-resolution power. However, it is worth mentioning that the proposed algorithm is more powerful to separate close layover. At the top the building, most reflections from roof and facade are overlaid. Comparing to SL1MMER, the proposed algorithm captures more reflection from the facade. This confirms the finding in Fig. \ref{fig:comp_PM7} that the proposed algorithm outperforms SL1MMER at low number of measurements. 


In terms of detection rate, the proposed algorithm is comparable to that of SL1MMER. Although there is no ground truth, we compare the agreement of the double scatterers detection of the both algorithms (shown in Table \ref{tab:n_scatterer}). For the whole test site, $38.97 \%$ and $37.76 \%$ of pixels are detected as double scatterers by the proposed algorithm and SL1MMER, respectively. $36.56\%$ of the pixels were detected as double scactterers by both algorithms. Only $2.4\%$ were only detected double scatterers by the proposed algorithm, and only $1.2\%$ were only detected by SL1MMER.
\begin{table}[h]
    \centering
    \begin{tabular}{p{0.1\textwidth}| p{0.1\textwidth} p{0.1\textwidth} p{0.1\textwidth}}
    \toprule
    \multirow{2}{*}{Algorithm}  & \multicolumn{3}{c}{Percentage of detection as} \\
     & \textcolor{red}{0} scatterer & \textcolor{red}{1} scatterer & \textcolor{red}{2} scatterers \\
     \midrule
     proposed & 30.71 $\%$ & 30.32 $\%$ & 38.97 $\%$   \\
     SL1MMER & 30.41 $\%$ & 31.83 $\%$ & 37.76 $\%$  \\
     \bottomrule
    \end{tabular}
    \caption{Percentage of scatterers detection for the two algorithms.}
    \label{tab:n_scatterer}
\end{table}




\section{Discussion}
\subsection{Analysis of computational complexity}
We assume $\mathcal{O}(1)$ to be the computational complexity of one multiplication. The computational complexity of the proposed algorithm, as well as the original ISTA, are mainly determined by $\mathcal{O}(K_s L^2)$, where $K_s$ is the number of layers or the number of iterations. For the proposed algorithm, $K_s$ is set as 12. Comparing to the original ISTA, which usually requires hundreds or even thousands of iteration, the computational efficiency of the proposed algorithm is two to three orders of magnitude better. Moreover, other efficient $L_1$ norm minimization solvers, such as FISTA \cite{FISTA}, ADMM \cite{admm}, RBPG \cite{Shi2018Fast}, usually need about 100 iterations to converge and achieve reasonable estimation accuracy. Comparing to those efficient solvers, the proposed algorithm is still about one order of magnitude more efficient. 

\begin{figure*}
    \centering
    \begin{minipage}[t]{0.49\linewidth}
    \includegraphics[width=0.98\textwidth]{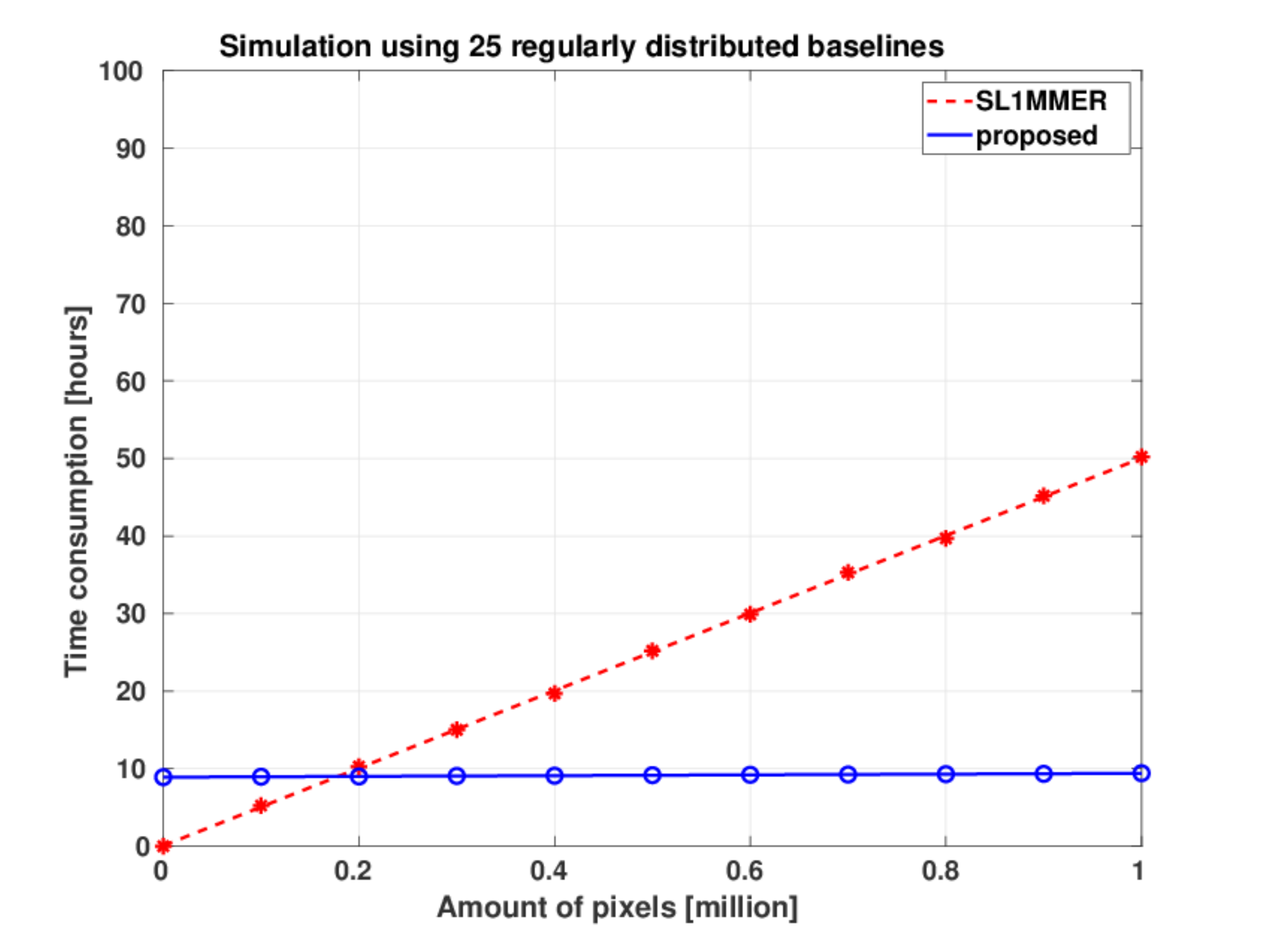}
    \caption*{(a)}
    \end{minipage}
    \begin{minipage}[t]{0.49\linewidth}
    \includegraphics[width=0.98\textwidth]{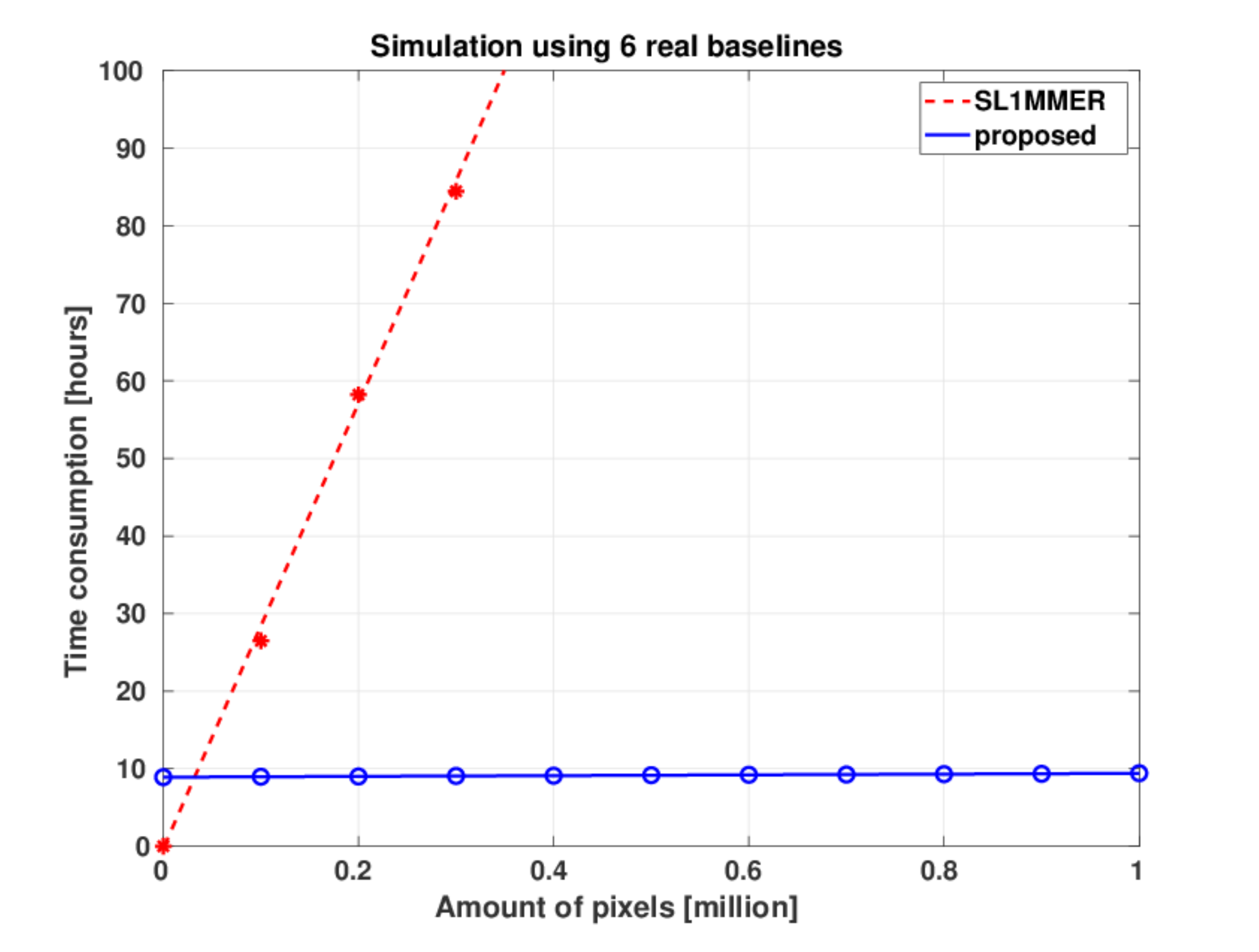}
    \caption*{(b)}
    \end{minipage}
    \caption{Comparison of time consumption between the proposed algorithm and SL1MMER. (a) on dataset simulated using 25 regularly distributed baselines at 6 dB, (b) on dataset simulated using 6 real baselines at 6 dB. The training time will be affected only by the size of training data and the number of training epochs we set. Different baseline configuration does not affect the training time. On the contrary, the time consumption of SL1MMER is strongly dependent on the number of baselines. When limited images are available, the time consumption of SL1MMER escalates with the increasing number of data, whereas the inference time of the trained $\boldsymbol{\gamma}$-Net is negligible. The proposed algorithm shows great computational efficiency in processing regular TomoSAR data, which usually contains tens of million pixels}
    \label{fig:time_consumption}
\end{figure*}

In our experiments, for a single dataset containing 0.2 million Monte Carlo trials simulated using the aforementioned setup, SL1MMER requires about 10 CPU hours for the TomoSAR processing. On the contrary, it takes only a few CPU minutes when a trained $\boldsymbol{\gamma}$-Net is employed, despite the fact that about 9 hours are required for training the model with a single NVIDIA RTX 2080 GPU. However, the fixed cost of model training diminishes when we further increase the amount of the data. Fig. \ref{fig:time_consumption} provides us an intuitive view of the time consumption of the two methods for TomoSAR processing. As we can see, the training procedure dominates the time consumption of the proposed algorithm and the increment of the amount of data will not burden the time consumption seriously. In the contrast, the time consumption of SL1MMER escalates with the increasing amount of data, especially when limited measurements are available. In real-world TomoSAR processing, the number of pixels is usually tens or even hundreds of million, thus blocking the application of SL1MMER or other second-order CS-based methods. The proposed algorithm is able to complete the processing, including the training procedure, within matters of hours. The great superiority of the proposed algorithm in computational efficiency makes large-scale super-resolving TomoSAR processing feasible and realizable.

In addition, it is worth mentioning that the proposed algorithm maintains the elevation estimation accuracy in the meanwhile. The proposed algorithm employed the neural network with special structure, which can be trained as a more general model and is more likely to reach the global minimum and achieve better results. A detailed investigation about how deep learning improves the estimation efficiency for TomoSAR inversion will be executed in our following study.

\subsection{Parameter selection}
\subsubsection*{\textbf{Step size in $\boldsymbol{\gamma}$-Net}}
As it is stated in the equation (\ref{eq:LISTA}), a manual selected step size is required for the initialization of the trainable weights in $\boldsymbol{\gamma}$-Net. To select the step size, the Lipschitz constant $L_s$ is required, which is the largest eigenvalue of $\mathbf{R^{H}R}$. Usually, a proper step size can be taken as $\frac{1}{L_s}$. In our experiments, we fix the step size as $\frac{1}{2 L_s}$ to guarantee the convergence of $\boldsymbol{\gamma}$-Net.

\subsubsection*{\textbf{Percentage $\rho$ in the support selection}}
An empirical formula is introduced in \cite{LISTA_cpss} to choose a proper percentage $\rho^i$ for the $i^{th}$ layer of $\boldsymbol{\gamma}$-Net for the support selection,
\begin{equation}
    \rho^i = max(p \cdot i, p_{max})
\end{equation}
where $p$ is a positive constant and $p_{max}$ is the upper bound of the percentage of the support cardinality. Both $p$ and $p_{max}$ can be selected using cross validation. From our experience, we fix the percentage $\rho_i$ as $5\%$ for all layers of $\boldsymbol{\gamma}$-Net, which leads to satisfactory performance for our application.

\subsection{Piecewise linear function}
The learning architecture of $\boldsymbol{\gamma}$-Net, where the output of the current layer is generated only directly from the output of the previous layer, leads to an error propagation phenomenon. Specifically, errors in the first few layers of $\boldsymbol{\gamma}$-Net will be propagated and further amplified in the following layers. Moreover, the most serious problem is that once useful information is discarded in the previous layers, it is no longer possible for the upcoming layers to utilize the discarded information. The conventional soft-thresholding function, simply prunes elements, whose magnitude is smaller than the threshold, to zero, which is very likely to discard information. The piecewise linear function is a smooth alternative of the soft-thresholding function. Instead of simply pruning elements with small magnitude, the piecewise linear function just further minifies them to maintain information as much as possible and execute the shrinkage step in the meanwhile, thus moderating the information loss caused by the learning architecture and improving the performance.

\subsection{Limitations of the proposed algorithm}
\subsubsection*{\textbf{\textit{Training time}}}
In the training procedure, there are $2 \times N L K + 5 \times K$ free trainable parameters. From our experience, about 2000 epochs are usually required for training the model, which takes about 9 hours when a single NVIDIA RTX 2080 GPU is employed. Due to the inevitable time consumption of the training, the proposed algorithm is not recommended for processing small datasets because it consumes less time when conventional CS-based methods are used.

Similar to all other deep learning based models, the training time could also become a burden when the task is to process many stacks with distinct baseline distributions. In our experiments, we simulated training data with the exact baselines as the test data. Ideally in real data processing, we shall train a separate model for each stack, which is very time consuming. However, our models shows moderate tolerance to the baseline discrepancies between the training and the testing data. This is elaborated in more detail in the next section.

\subsubsection*{\textbf{\textit{Baseline perturbation}}}
The biggest challenge to our deep learning model is the baseline discrepancies between the training and the testing data, because the baseline distribution are rather unique for each SAR interferometric stack. As a preliminary study, we test the proposed algorithm on testing data  with slight baseline perturbation and find that the slight perturbation do not degrade the performance significantly. To be specific, we add random perturbation uniformly distributed in the range  $[-10m,10m]$, i.e. about $15\%$ of the baseline standard deviation, to the 25 baselines. Applying the pre-trained $\boldsymbol{\gamma}$-Net on the data with baselines perturbation shows  that the effective detection rate decreases only $3\%$ to $5\%$ and the estimation accuracy as well as the bias retain nearly the same. This shows a good transferability of our trained model. However, further study is required to guarantee the performance of the proposed algorithm for large scale processing.

\subsubsection*{\textbf{\textit{Application to more complex scenarios}}}
When the proposed algorithm is applied to more complex scenarios, i.e. more than 2 scatterers are overlaid in a single resolution unit, we are not capable of detecting and separating all of them. We tested the proposed algorithm in the three-scatterer case and found that the proposed algorithm tends to detect overlaid triple scatterers as double scatterers locating between the ground truth. Due to the fact that triple scatterers are not considered and covered in the training phase, the poor performance in coping with triple scatterers is explainable. From our perspective, the solution to this problem can be two-fold. First, we can enrich the training data by introducing samples containing more scatterers. Second, we can view samples containing more than two scatterers as out-of-distribution samples, since in real-world processing only a tiny minority of pixels contain more than two scatterers. We can then use Dirichlet Prior Network (DPN) \cite{DPN1} \cite{DPN2} to detect these out-of-distribution data and solve it using CS-solvers.

\section{Conclusion}
In this paper, an advanced super-resolution TomoSAR inversion approach based on deep learning is proposed. We improved the complex-valued learned ISTA algorithm and proposed $\boldsymbol{\gamma}$-Net by applying weight coupling structure, introducing support selection and employing the piecewise linear function instead of soft-thresholding. Experiments show that the proposed algorithm is capable of solving the $L_2$-$L_1$ mixed norm minimization efficiently. Rigorous evaluation shows that the proposed approach is able to deliver competitive performance to the state of the art in terms of the super-resolution capability and elevation estimation accuracy. 
This paper opens a perspective on super-resolving TomoSAR inversion via deep learning and shows great potential of applying deep learning to solve other sparse reconstruction problems. In the future, we aim to extend the deep learning based approach to higher dimensional spectral estimation problems, especially to differential TomoSAR reconstruction. Moreover, we will further exploit the power of deep learning to improve the performance, e.g. introducing long short term memory (LSTM) unit to $\boldsymbol{\gamma}$-Net to make use of historic information.


%



\ifCLASSOPTIONcaptionsoff
  \newpage
\fi



%
\bibliographystyle{IEEEtran}
\bibliography{ref}



%
\begin{IEEEbiography}[{\includegraphics[width=1in,height=1.25in,clip,keepaspectratio]{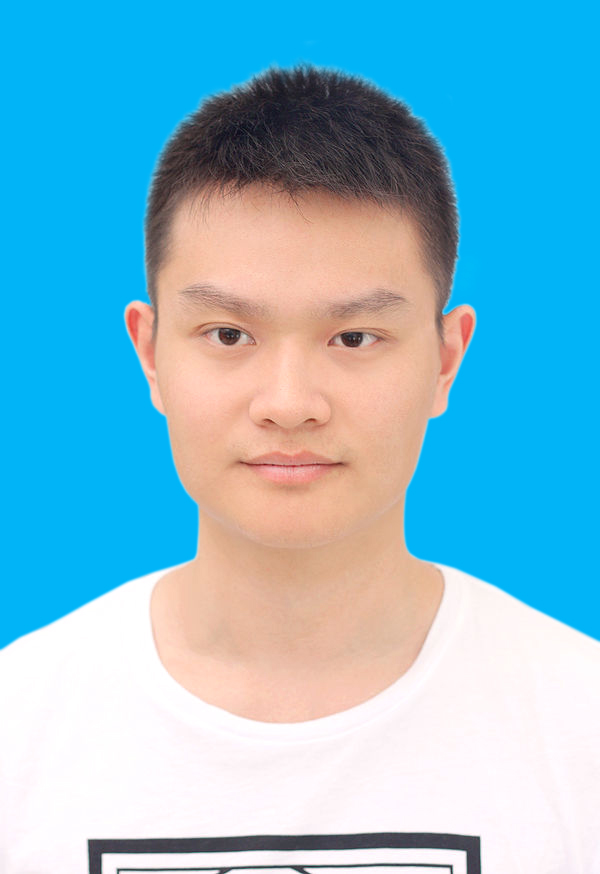}}]{Kun Qian}
received double B.Sc. degree in Remote Sensing and Information Engineering from Wuhan University, Wuhan, China and Aerospace Engineering and Geodesy from University of Stuttgart, Stuttgart, Germany, respectively, in 2016, and M.Sc. degree in Aerospace Engineering and Geodesy from University of Stuttgart, Stuttgart, Germany in 2018. He is pursuing the Ph.D. degree with Data Science in Earth Observation, Technical Unversity of Munich, Munich, Germany. His research focus includes data-driven methods, deep unfolding algorithms and their application in multi-baseline SAR tomography.
\end{IEEEbiography}

\begin{IEEEbiography}[{\includegraphics[width=1in,height=1.25in,clip,keepaspectratio]{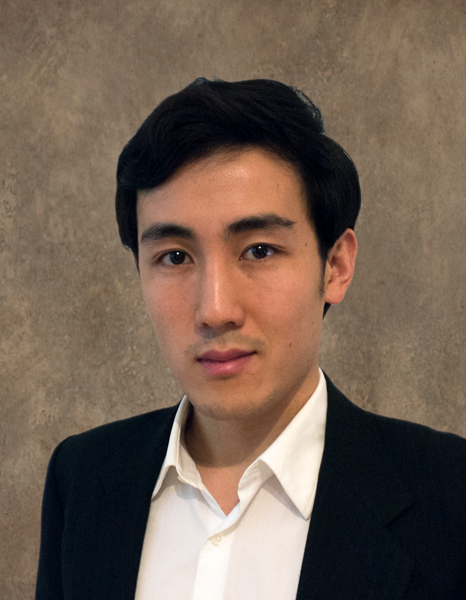}}]{Yuanyuan Wang}
received the B.Eng. degree (Hons.) in Electrical Engineering from The Hong Kong Polytechnic University, Hong Kong, China in 2008, and the M.Sc. and Dr.-Ing. degree from the Technical University of Munich, Munich, Germany, in 2010 and 2015, respectively. In June and July 2014, he was a Guest Scientist with the Institute of Visual Computing, ETH Zürich, Zürich, Switzerland. He is currently with the Department of EO Data Science, in the Remote Sensing Technology Institute of the German Aerospace Center, where he leads the working group Big SAR Data. His research interests include optimal and robust parameters estimation in multibaseline InSAR techniques, multisensor fusion algorithms of synthetic aperture radar and optical data, nonlinear optimization with complex numbers, machine learning in SAR, and high-performance computing for big data. Dr. Wang was one of the best reviewers of the IEEE TRANSACTIONS ON GEOSCIENCE AND REMOTE SENSING in 2016. He is a Member of the IEEE.
\end{IEEEbiography}

\begin{IEEEbiography}[{\includegraphics[width=1in,height=1.25in,clip,keepaspectratio]{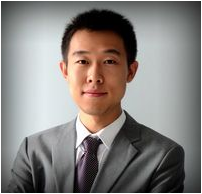}}]{Yilei Shi} (Member, IEEE) received the Dipl.-Ing. degree in mechanical engineering and the Dr.-Ing. degree in signal processing from Technische Universität München (TUM), Munich, Germany, in 2010 and 2019, respectively. He is a Senior Scientist with the Chair of Remote Sensing Technology, TUM. His research interests include fast solver and parallel computing for large scale problems, high-performance computing and computational intelligence, advanced methods on SAR and InSAR processing, machine learning and deep learning for variety of data sources, such as SAR, optical images, and medical images, and PDE-related numerical modeling and computing.
\end{IEEEbiography}

\begin{IEEEbiography}[{\includegraphics[width=1in,height=1.25in,clip,keepaspectratio]{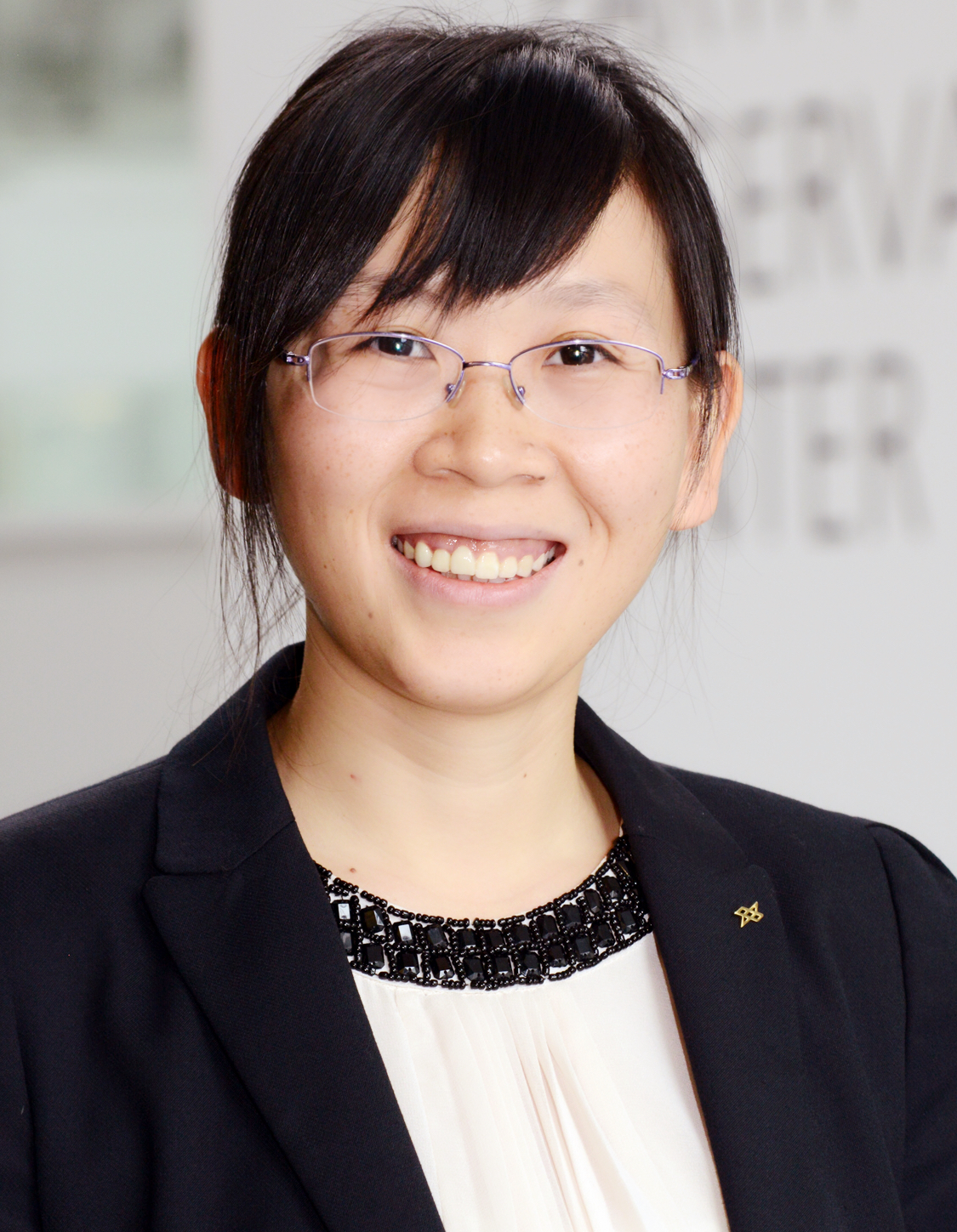}}]{Xiao Xiang Zhu}(S'10--M'12--SM'14--F'21) received the Master (M.Sc.) degree, her doctor of engineering (Dr.-Ing.) degree and her “Habilitation” in the field of signal processing from Technical University of Munich (TUM), Munich, Germany, in 2008, 2011 and 2013, respectively.
\par
She is currently the Professor for Data Science in Earth Observation (former: Signal Processing in Earth Observation) at Technical University of Munich (TUM) and the Head of the Department ``EO Data Science'' at the Remote Sensing Technology Institute, German Aerospace Center (DLR). Since 2019, Zhu is a co-coordinator of the Munich Data Science Research School (www.mu-ds.de). Since 2019 She also heads the Helmholtz Artificial Intelligence -- Research Field ``Aeronautics, Space and Transport". Since May 2020, she is the director of the international future AI lab "AI4EO -- Artificial Intelligence for Earth Observation: Reasoning, Uncertainties, Ethics and Beyond", Munich, Germany. Since October 2020, she also serves as a co-director of the Munich Data Science Institute (MDSI), TUM. Prof. Zhu was a guest scientist or visiting professor at the Italian National Research Council (CNR-IREA), Naples, Italy, Fudan University, Shanghai, China, the University  of Tokyo, Tokyo, Japan and University of California, Los Angeles, United States in 2009, 2014, 2015 and 2016, respectively. She is currently a visiting AI professor at ESA's Phi-lab. Her main research interests are remote sensing and Earth observation, signal processing, machine learning and data science, with a special application focus on global urban mapping.

Dr. Zhu is a member of young academy (Junge Akademie/Junges Kolleg) at the Berlin-Brandenburg Academy of Sciences and Humanities and the German National  Academy of Sciences Leopoldina and the Bavarian Academy of Sciences and Humanities. She serves in the scientific advisory board in several research organizations, among others the German Research Center for Geosciences (GFZ) and Potsdam Institute for Climate Impact Research (PIK). She is an associate Editor of IEEE Transactions on Geoscience and Remote Sensing and serves as the area editor responsible for special issues of IEEE Signal Processing Magazine. She is a Fellow of IEEE.
\end{IEEEbiography}




\end{document}